\documentclass[Afour,sageh,times]{sagej}

\usepackage{moreverb,url}
\usepackage{float} 
\usepackage{amsmath}
\usepackage{siunitx}
\usepackage{graphicx}
\usepackage{todonotes}
\usepackage{subcaption}  %
\usepackage{accents} %
\usepackage{braket}
\usepackage{bm}

\def\tsc#1{\csdef{#1}{\textsc{\lowercase{#1}}\xspace}}
\tsc{WGM}
\tsc{QE}

\usepackage{dblfloatfix} %
\usepackage[colorlinks,bookmarksopen,bookmarksnumbered,citecolor=red,urlcolor=red]{hyperref}

\newcommand\BibTeX{{\rmfamily B\kern-.05em \textsc{i\kern-.025em b}\kern-.08em
T\kern-.1667em\lower.7ex\hbox{E}\kern-.125emX}}

\def\volumeyear{2016}

\begin{document}

\runninghead{Hauck and Turkalj}

\title{Classical Tensor Train and Quantum Fourier Transform Approaches for Large-Scale Carr-Madan Option Pricing}

\author{Sascha H. Hauck\affilnum{1, 3} and Ivica Turkalj\affilnum{2}}

\affiliation{\affilnum{1}Fraunhofer ITWM, Department of Material Simulation,\\ Kaiserslautern, 67663, Germany\\
\affilnum{2}Fraunhofer ITWM, Department of Financial Mathematics, \\Kaiserslautern, 67663, Germany\\
\affilnum{3}University of Kaiserslautern-Landau, Chair for Scientific Computing, \\Kaiserslautern, 67663, Germany
}

\corrauth{Sascha H. Hauck}
\email{sascha.hannes.hauck@itwm.fraunhofer.de}

\begin{abstract}
Fourier-based methods are widely used for pricing European options when the characteristic function of the underlying is available. However, for increasingly fine discretizations, the memory requirements and runtime of the classical Fourier transform become a computational bottleneck.

We address this limitation by reformulating the Carr-Madan pricing algorithm using tensor trains. Specifically, we employ the Superfast Fourier Transform (SFFT), a compressed tensor train representation of the Quantum Fourier Transform (QFT), without ever explicitly constructing exponentially large vectors or Fourier operators. This formulation also enables a direct comparison with a QFT-based implementation of the pricing algorithm on quantum hardware.

Numerical experiments for European call options under the Black-Scholes and Heston models, including geometric baskets of five assets, demonstrate that the SFFT maintains pricing accuracy while reducing memory requirements by almost five orders of magnitude and achieving subexponential scaling compared with conventional FFT-based pricing. For a genuinely multidimensional worst-of call, the SFFT handles up to sixteen assets, corresponding to $2^{96}$ grid points, with less than $11$\,MB of memory. Although the QFT approach also exhibits improved scalability, we find no fundamental advantage in employing quantum hardware for this problem.
\end{abstract}

\maketitle

\section{Introduction}
Financial options are derivative contracts whose value depends on the future
evolution of an underlying asset. A European call option, for example, gives
its holder the right, but not the obligation, to buy the asset at a fixed
strike price $K$ at maturity $T$. Under the absence of arbitrage, the
price of such a contract can be represented as the discounted expectation of
its payoff under a risk-neutral probability measure \citep{harrisonPliska1981}.

In general asset-price models, closed form option prices are
often unavailable even when the distributional structure of the model is
known analytically. This situation arises, for example, in models with jumps,
stochastic volatility, or other non-Gaussian return dynamics
\citep{merton1976,heston1993}. However, in many such
models the characteristic function of the log-price
is available in closed form
\citep{heston1993,carrMadan1999,lee2004,duffiePanSingleton2000}. 
Fourier pricing methods exploit this fact by
rewriting option prices in terms of Fourier transforms. Instead of computing
the risk-neutral expectation directly, the method
works in frequency space, where the characteristic function provides a compact
description of the distribution of the log-price.

For European options, the Fourier representation can be discretized on a
finite grid. After introducing a damping parameter to ensure
integrability of the transformed payoff, the option value can be recovered
from an inverse Fourier transform \citep{carrMadan1999,lee2004}. 
Numerically, this leads to a discrete
Fourier transform over a grid of frequencies and log-strikes. The resulting
structure is computationally attractive because a complete set of option
prices on a strike grid can be obtained simultaneously, rather than pricing
each strike independently.

Since the introduction of the Carr-Madan FFT approach, a variety of
Fourier-based pricing techniques have been developed. A prominent alternative
is the Fourier-cosine method, which directly exploits the characteristic
function through a cosine-series expansion and can achieve rapid, and under
suitable regularity conditions exponential, convergence
\citep{fangOosterlee2008}. Recent work has further established rigorous
error-control and parameter-selection criteria for the cosine method, while
clarifying its convergence properties for models with different tail
behavior \citep{junike2024}.

Recent years have witnessed increasing interest in tensor network methods for
computational finance. Their ability to represent certain exponentially large
arrays through low-rank decompositions has motivated applications to high-dimensional
option pricing, where tensor network representations can mitigate the exponential
growth of computational complexity with the number of assets \citep{kastoryano22}.
Related work has used tensor train approximations of Fourier-based pricing
functions to accelerate parameter exploration, pricing, and Greek computation
\citep{sakurai25greeks,sakurai25learning,glau20}, demonstrating the
potential of tensor network methods in the field.

Most existing tensor network approaches exploit low-rank structure in the tensors arising from the pricing problem, for example through representations of characteristic, payoff or pricing function, model parameters, or sensitivity surfaces.
In contrast, the tensor-product structure of the Fourier transformation operator itself has received comparatively little attention.
At the same time, quantum-computing approaches for option pricing  naturally exploit this structure through the Quantum Fourier Transform (QFT) \citep{stamatopoulos2020,ewen2024}.
This raises the question of whether the analogous tensor-product structure of the QFT can be exploited efficiently on classical hardware by means of tensor networks.

In this work, we address this question by developing a tensor train formulation
of the Carr-Madan Fourier pricing algorithm based on the Superfast Fourier
Transform (SFFT), a compressed tensor network realization of the
Quantum Fourier Transform.
The SFFT permits Fourier transformations to be efficiently performed directly on tensor trains.
Consequently, the pricing algorithm can operate on exponentially large
Fourier grids without explicitly constructing the corresponding dense vectors
whenever the relevant quantities admit sufficiently low TT ranks.
This establishes a connection between classical FFT-based pricing, tensor network algorithms, and quantum computing approaches.

The contribution of this work is threefold. First, we formulate the Carr-Madan
Fourier pricing procedure within a tensor train framework by employing the Superfast Fourier Transform. Second, we investigate the relationship between the classical FFT, the QFT evaluated on quantum hardware and simulators, and the SFFT by implementing and comparing all three approaches within a common option-pricing framework.
Third, we conduct extensive numerical experiments to evaluate pricing
accuracy, memory requirements, computational scaling, and hardware
execution characteristics. These experiments cover single-asset and
geometric-basket call options under the Black-Scholes and Heston
models. In addition, we investigate the scalability of the
SFFT-based approach with respect to the number of assets using
worst-of call options.

The remainder of this paper is organized as follows.
We first introduce the classical FFT-based option pricing framework and derive the Carr-Madan discretization used throughout this work, together with its multidimensional
extension to worst-of calls, followed by the corresponding QFT-based formulation and quantum pricing algorithm.
We then introduce the tensor train representation and the proposed SFFT-based pricing approach.
Afterwards, numerical experiments are conducted to compare the classical FFT-based and tensor-network methods with their quantum counterpart on quantum simulators and quantum hardware.
Finally, we summarize our findings and discuss possible improvements and directions for future work.

\section{Option Pricing}
\label{sec:general_option_pricing}
As noted in the Introduction, an option is a financial instrument whose payoff depends on the stochastic evolution of the prices of one or more underlying assets. Under the standard no-arbitrage assumptions, the value of a replicable option can be expressed as the discounted expectation of its future payoff under an appropriate risk-neutral probability measure \citep{baxterRennie1996, shreve2004}. In many asset-pricing models, however, the probability density of the underlying asset prices is not available in closed form. Consequently, the expectation defining the option value may not be analytically tractable and may require numerical approximation. In the following, we investigate this issue in more detail and discuss possible approaches for overcoming the resulting computational difficulties.

\subsection{Single-asset Black-Scholes Model}
\label{subsec:univariate_black_scholes}
Let $S_t>0$ denote the price of an underlying asset at time
$t\geq 0$. The evolution of the stochastic process $(S_t)_{t\geq 0}$
is specified by a stochastic differential equation.
For example, in the Black-Scholes model \citep{blackScholes1973, merton1973}, $S_t$ denotes the price of a stock whose dynamics are governed by the equation 
\begin{align*}
    \mathrm{d}S_t = S_t r_{\mathrm{f}}\mathrm{d}t + S_t \sigma\mathrm{d}W_t^{\mathbb{Q}},
\end{align*}
where $r_{\mathrm{f}}\in\mathbb{R}$ is the constant risk-free interest rate,
$\sigma>0$ is the volatility, and $W^{\mathbb{Q}}$ is a standard
Brownian motion under the risk-neutral measure $\mathbb{Q}$.
Using Itô-calculus, it can be shown that the stochastic process solving above equation is of the form
\begin{align*}
    S_t = S_0 e^{\sigma W_t^{\mathbb{Q}} + (r_{\mathrm{f}} - \sigma^2/2)t},
\end{align*}
for every $t$,
where $S_0>0$ is the deterministic initial price of the stock.

In general, an option's payoff may depend on the entire path $(S_t)_{0\leq t\leq T}$ of the underlying asset.
Here, we focus on path-independent European options, whose payoff at maturity $T$ is modeled by the random variable $f(S_T)$, where $f\colon \mathbb{R} \longrightarrow \mathbb{R}$.
The function $f$ is also referred to as the \emph{payoff function}. For example, the payoff of a European call option is of the form
\begin{align*}
    f(x) = \max \{0, x-K\},
\end{align*}
where the parameter $K > 0$ is called the strike of the option.
Under the standard no-arbitrage assumptions and in a complete market, the time-zero value of a replicable European claim is given by its discounted risk-neutral expected payoff \citep{merton1973,baxterRennie1996},
\begin{align*}
   e^{-r_{\mathrm{f}}T} \mathbb{E}^{\mathbb{Q}}\left[f(S_T)\right]
\end{align*}
where $\mathbb{E}^{\mathbb{Q}}$ is the expected value. 
For a non-dividend-paying asset, the time-zero price of a European call
option in the Black-Scholes model has a standard closed-form expression
\citep{shreve2004}.

The corresponding characteristic function of the terminal log-price is
\begin{align}
    \label{eq:univariate_bs_cf}
    \phi_{\mathrm{BS}}(u;T)
    &= \mathbb{E}^{\mathbb{Q}}\left[e^{iu\log S_T}\right]\nonumber\\
    &= \exp\Bigl(iu[\log S_0+(r_{\mathrm{f}}-\sigma^2/2)T]\nonumber\\
    &\qquad-\sigma^2Tu^2/2\Bigr).
\end{align}

In many stochastic asset-pricing models, the stochastic differential equation governing $S_t$ does not admit an explicit solution, or the transition density of $S_t$ is not available in closed form. Consequently, the risk-neutral expectation defining the option price may not be analytically tractable. In those case, numerical valuation is applied. Depending on the structure of the model, option prices can be approximated using Monte Carlo simulation \citep{boyle1977}, numerical solutions of PDEs, lattice methods, numerical quadrature, or Fourier-transform techniques. In the following, we focus on models for which the characteristic function of the log-price is known explicitly, allowing the pricing expectation to be evaluated efficiently.

\subsection{Multi-asset Black-Scholes Model}
\label{subsec:multivariate_black_scholes}
The Black-Scholes model extends to $d$ assets by replacing the single
Brownian motion with correlated Brownian motions
\citep{shreve2004}. For $j=1,\ldots,d$, let
$W_j^{\mathbb{Q}}=(W_{j,t}^{\mathbb{Q}})_{t\geq 0}$
denote the Brownian process driving asset $j$. The risk-neutral dynamics are
\begin{align}
    \label{eq:multivariate_bs_dynamics}
    \mathrm{d}S_{j,t}
    = (r_{\mathrm{f}}-q_j)S_{j,t}\,\mathrm{d}t
      + \sigma_j S_{j,t}\,\mathrm{d}W_{j,t}^{\mathbb{Q}},
\end{align}
where $S_{j,0}>0$, $\sigma_j\geq 0$, and $q_j$ is a constant dividend
yield. The quadratic covariation of the Brownian processes, evaluated
at time $t$, satisfies
\begin{align*}
    \mathrm{d}\langle W_i^{\mathbb{Q}},W_j^{\mathbb{Q}}\rangle_t
    =R_{ij}\,\mathrm{d}t,
\end{align*}
where $R$ is a symmetric positive-semidefinite
correlation matrix with unit diagonal. With
$\Sigma_{ij}=\sigma_i R_{ij}\sigma_j$, the terminal log-price vector
$\bm{s}_T=(\log S_{1,T},\ldots,\log S_{d,T})^\top$ is Gaussian:
\begin{align*}
    \bm{s}_T &\sim \mathcal{N}(\bm{m}_T,T\Sigma),\\
    m_{j,T} &= \log S_{j,0}
        +(r_{\mathrm{f}}-q_j-\sigma_j^2/2)T.
\end{align*}
Consequently, its joint characteristic function is
\begin{align}
    \label{eq:multivariate_bs_cf}
    \phi_{\mathrm{mBS}}(\bm{u};T)
    &= \mathbb{E}^{\mathbb{Q}}
       \left[e^{i\bm{u}^{\top}\bm{s}_T}\right]\nonumber\\
    &= \exp\left(i\bm{u}^{\top}\bm{m}_T
       -\frac{T}{2}\bm{u}^{\top}\Sigma\bm{u}\right).
\end{align}

We consider European calls on a weighted geometric basket,
\begin{align}
    \label{eq:geometric_basket}
    B_t &= \prod_{j=1}^{d}S_{j,t}^{w_j},
    \qquad w_j\geq 0,\quad \sum_{j=1}^{d}w_j=1,\\
    \label{eq:geometric_basket_price}
    C_B(K) &= e^{-r_{\mathrm{f}}T}
       \mathbb{E}^{\mathbb{Q}}\left[(B_T-K)^+\right],
\end{align}
where $x^+=\max\{0,x\}$. Since
$\log B_T=\bm{w}^{\top}\bm{s}_T$, the basket is itself lognormal.
Its distribution is described by the effective parameters
\begin{align}
    \label{eq:bs_basket_parameters}
    B_0 &= \prod_{j=1}^{d}S_{j,0}^{w_j},\qquad
    \sigma_B^2=\bm{w}^{\top}\Sigma\bm{w},\nonumber\\
    q_B &= \sum_{j=1}^{d}w_jq_j
       +\frac{1}{2}\left(\sum_{j=1}^{d}w_j\sigma_j^2-\sigma_B^2\right).
\end{align}
In particular, $q_B$ is generally nonzero even when all individual
dividend yields vanish. This correction accounts for the difference
between the weighted individual variances and the basket variance.
The resulting analytical call price is
\begin{align}
    \label{eq:analytic_bs_basket}
    C_B(K) &= B_0e^{-q_BT}\Phi(d_{1,B})
       -Ke^{-r_{\mathrm{f}}T}\Phi(d_{2,B}),\\
    d_{1,B} &=
       \frac{\log(B_0/K)
       +(r_{\mathrm{f}}-q_B+\sigma_B^2/2)T}
       {\sigma_B\sqrt{T}},\nonumber\\
    d_{2,B} &= d_{1,B}-\sigma_B\sqrt{T},\nonumber
\end{align}
where $\Phi$ is the cumulative distribution function of the standard
normal distribution.
For $\sigma_B=0$, the price is the discounted deterministic payoff,
obtained by continuity. Equivalently, Fourier pricing uses the scalar
characteristic function
$\phi_B(u;T)=\phi_{\mathrm{mBS}}(u\bm{w};T)$.

\subsection{Single-asset Heston Model}
\label{subsec:univariate_heston}
The Heston model \citep{heston1993} replaces constant variance by a
nonnegative stochastic variance process $V_t$. Under a chosen
risk-neutral measure $\mathbb{Q}$, let
$W^{\mathbb{Q}}=(W_t^{\mathbb{Q}})_{t\geq 0}$ and
$Z^{\mathbb{Q}}=(Z_t^{\mathbb{Q}})_{t\geq 0}$
denote the Brownian processes driving the asset price and variance,
respectively. The asset and variance satisfy
\begin{align}
    \label{eq:heston_asset_dynamics}
    \mathrm{d}S_t
      &= (r_{\mathrm{f}}-q)S_t\,\mathrm{d}t
         +S_t\sqrt{V_t}\,\mathrm{d}W_t^{\mathbb{Q}},\\
    \label{eq:heston_variance_dynamics}
    \mathrm{d}V_t
      &= \kappa(\theta-V_t)\,\mathrm{d}t
         +\xi\sqrt{V_t}\,\mathrm{d}Z_t^{\mathbb{Q}},
\end{align}
with $S_0>0$ and $V_0=v_0\geq 0$. The quadratic covariation of
the Brownian processes, evaluated at time $t$, satisfies
\begin{align*}
    \mathrm{d}\langle W^{\mathbb{Q}},Z^{\mathbb{Q}}\rangle_t
    =\rho\,\mathrm{d}t,\qquad |\rho|\leq 1.
\end{align*}
Here $q$ is the dividend yield,
$\kappa>0$ is the mean-reversion rate, $\theta\geq 0$ is the
long-run variance, and $\xi\geq 0$ is the volatility of variance,
which is distinct from the asset volatility $\sqrt{V_t}$.
For $v_0>0$, the Feller condition $2\kappa\theta\geq\xi^2$ is
sufficient for strict positivity of $V_t$. It is not required for
nonnegativity or for the characteristic-function representation.
The model is generally incomplete, so no-arbitrage alone does not
select a unique pricing measure. The parameters above are specified
under $\mathbb{Q}$ and therefore already incorporate the choice of
variance risk premium.

The log-price satisfies
\begin{align*}
    \mathrm{d}\log S_t
    = (r_{\mathrm{f}}-q-V_t/2)\,\mathrm{d}t
      +\sqrt{V_t}\,\mathrm{d}W_t^{\mathbb{Q}}.
\end{align*}
Although the European call price is not given by the elementary
Black-Scholes formula, the characteristic function of the terminal
log-price is available in exponential-affine form
\citep{heston1993,duffiePanSingleton2000}:
\begin{align}
    \label{eq:heston_cf}
    \phi_{\mathrm{H}}(u;T)
    &= \mathbb{E}^{\mathbb{Q}}\left[e^{iu\log S_T}\right]\nonumber\\
    &= \exp\Bigl(iu[\log S_0+(r_{\mathrm{f}}-q)T]\nonumber\\
    &\qquad+A(u,T)+D(u,T)v_0\Bigr).
\end{align}
For $\xi>0$, define
\begin{align*}
    \beta(u) &= \kappa-i\rho\xi u,\\
    \delta(u) &= \sqrt{\beta(u)^2+\xi^2(u^2+iu)},\\
    g(u) &= \frac{\beta(u)-\delta(u)}{\beta(u)+\delta(u)}.
\end{align*}
The affine coefficients are
\begin{align}
    A(u,T)
    &= \frac{\kappa\theta}{\xi^2}
       \biggl[(\beta(u)-\delta(u))T\nonumber\\
    &\qquad-2\log\left(
       \frac{1-g(u)e^{-\delta(u)T}}{1-g(u)}
       \right)\biggr],\label{eq:heston_A}\\
    D(u,T)
    &= \frac{\beta(u)-\delta(u)}{\xi^2}
       \frac{1-e^{-\delta(u)T}}{1-g(u)e^{-\delta(u)T}}.
       \label{eq:heston_D}
\end{align}
The square root is chosen with nonnegative real part and the logarithm
by continuous continuation from $T=0$. Complex arguments are admissible only where
the corresponding exponential moment is finite.
For $\xi=0$, the variance is deterministic, with integrated variance
\begin{align*}
    I_T=\int_0^T V_t\,\mathrm{d}t
       =\theta T+(v_0-\theta)\frac{1-e^{-\kappa T}}{\kappa}.
\end{align*}
In this case, the exponent contribution $A(u,T)+D(u,T)v_0$ is
$-(u^2+iu)I_T/2$. If additionally $v_0=\theta=\sigma^2$, the
Black-Scholes model with dividend yield $q$ is recovered.
In the stochastic-variance case, the call price is evaluated through
Fourier inversion of \eqref{eq:heston_cf}.

\subsection{Multi-asset Heston Model with Common Variance}
\label{subsec:multivariate_heston}
The multivariate Heston specification considered here couples $d$
assets to the same variance process \eqref{eq:heston_variance_dynamics}.
For $j=1,\ldots,d$, let
$W_j^{\mathbb{Q}}=(W_{j,t}^{\mathbb{Q}})_{t\geq 0}$
denote the Brownian process driving asset $j$. The asset-price dynamics are
\begin{align}
    \label{eq:multivariate_heston_dynamics}
    \mathrm{d}S_{j,t}
    = (r_{\mathrm{f}}-q_j)S_{j,t}\,\mathrm{d}t
      +a_j S_{j,t}\sqrt{V_t}\,\mathrm{d}W_{j,t}^{\mathbb{Q}},
\end{align}
where $a_j\geq 0$ is the volatility loading of asset $j$.
The common variance is driven by
$Z^{\mathbb{Q}}=(Z_t^{\mathbb{Q}})_{t\geq 0}$.
The quadratic covariations of the Brownian processes, evaluated
at time $t$, satisfy
\begin{align*}
    \mathrm{d}\langle W_i^{\mathbb{Q}},W_j^{\mathbb{Q}}\rangle_t
    &=R_{ij}\,\mathrm{d}t,\\
    \mathrm{d}\langle W_j^{\mathbb{Q}},Z^{\mathbb{Q}}\rangle_t
    &=\rho_j\,\mathrm{d}t,
\end{align*}
where $R_{ij}$ are the correlations between asset drivers and $\rho_j$
are their correlations with the common variance driver. A consistent
joint Brownian motion requires the full correlation matrix
\begin{align}
    \label{eq:heston_joint_correlation}
    \begin{pmatrix}
        R & \bm{\rho}\\
        \bm{\rho}^{\top} & 1
    \end{pmatrix}
    \succeq 0.
\end{align}
Valid pairwise correlations alone are not sufficient (we consider a
single-common-variance model, not a specification with separate
correlated variance processes for each asset).

Let $\Gamma_{ij}=a_iR_{ij}a_j$, and define
\begin{align}
    \label{eq:multivariate_heston_coefficients}
    h(\bm{u}) &= \bm{u}^{\top}\Gamma\bm{u}
                  +i\sum_{j=1}^{d}u_ja_j^2,\nonumber\\
    \ell(\bm{u}) &= \sum_{j=1}^{d}u_ja_j\rho_j.
\end{align}
The joint log-price characteristic function is again
exponential-affine:
\begin{align}
    \label{eq:multivariate_heston_cf}
    \phi_{\mathrm{mH}}(\bm{u};T)
    &= \mathbb{E}^{\mathbb{Q}}
       \left[e^{i\bm{u}^{\top}\bm{s}_T}\right]\nonumber\\
    &= \exp\Biggl(i\sum_{j=1}^{d}u_j
       [\log S_{j,0}+(r_{\mathrm{f}}-q_j)T]\nonumber\\
    &\qquad+A_{\mathrm{mH}}(\bm{u},T)
       +D_{\mathrm{mH}}(\bm{u},T)v_0\Biggr).
\end{align}
The coefficients $A_{\mathrm{mH}}$ and $D_{\mathrm{mH}}$ are given by
\eqref{eq:heston_A}-\eqref{eq:heston_D}, with the substitutions
\begin{align*}
    \beta(u)&\longmapsto\kappa-i\xi\ell(\bm{u}),\\
    \delta(u)&\longmapsto
      \sqrt{(\kappa-i\xi\ell(\bm{u}))^2+\xi^2h(\bm{u})},
\end{align*}
and with $g$ formed from these substituted quantities. This follows
from the affine structure of the common variance factor
\citep{duffiePanSingleton2000}. For $\xi=0$, the variance contribution
reduces to $-h(\bm{u})I_T/2$. Taking $d=1$ and $a_1=1$
recovers the univariate Heston model.

For the geometric basket in \eqref{eq:geometric_basket}, define
\begin{align}
    \label{eq:heston_basket_parameters}
    \bar q &= \sum_{j=1}^{d}w_jq_j,\qquad
    \lambda_B=\sum_{j=1}^{d}w_ja_j^2,\nonumber\\
    \nu_B &= \bm{w}^{\top}\Gamma\bm{w},\qquad
    \zeta_B=\sum_{j=1}^{d}w_ja_j\rho_j.
\end{align}
The log-basket dynamics are
\begin{align*}
    \mathrm{d}\log B_t
    &= (r_{\mathrm{f}}-\bar q-\lambda_B V_t/2)\,\mathrm{d}t\\
    &\quad+\sqrt{V_t}\sum_{j=1}^{d}w_ja_j\,
       \mathrm{d}W_{j,t}^{\mathbb{Q}}.
\end{align*}
The instantaneous variance of $\log B_t$ is $\nu_BV_t$, whereas
its log-drift contains $\lambda_BV_t$. Thus, unlike the
Black-Scholes basket, this basket cannot in general be represented
as an ordinary scalar Heston asset with a constant effective dividend
yield: the corresponding drift correction is
$(\lambda_B-\nu_B)V_t/2$ and is stochastic.
Nevertheless, its scalar characteristic function follows exactly by
projection,
\begin{align}
    \label{eq:heston_basket_cf}
    \phi_B(u;T)&=\phi_{\mathrm{mH}}(u\bm{w};T),\nonumber\\
    h(u\bm{w})&=\nu_Bu^2+i\lambda_Bu,\qquad
    \ell(u\bm{w})=\zeta_Bu.
\end{align}
The call price is still given by \eqref{eq:geometric_basket_price}.
For both multivariate models, the geometric payoff therefore permits
a one-dimensional Fourier inversion, rather than a $d$-dimensional
transform. This reduction does not apply to general multivariate
payoffs, such as calls on arithmetic baskets.

\subsection{European Option Pricing Based on Discrete Fourier Transform}
We employ the damped Fourier-transform formulation of \citet{carrMadan1999}
to derive a common pricing representation for all four models introduced
above. The transformation of the European call payoff and the subsequent
discrete Fourier transform (DFT) construction are shared. The
asset-price dynamics enter through the characteristic function of the
logarithm of a positive scalar underlying. The same derivation therefore
applies to both single-asset and geometric-basket calls as well as the worst-of calls.

A variety of Fourier-based option-pricing methods have been proposed in the literature. These schemes exploit analytical relationships between option prices and the characteristic function of the risk-neutral distribution of the underlying log-price \citep{heston1993,carrMadan1999,lee2004,fangOosterlee2008}. The Carr-Madan approach is particularly well suited to FFT-based evaluation because it applies an exponential damping factor to the call-price function, ensuring that its Fourier transform is well defined and producing a discretization with the structure of a discrete Fourier transform (DFT). In the following, we briefly derive this representation and its numerical implementation.

Let $T>0$ denote the maturity, let $K>0$ be the strike, and write
$k=\log K$. We use $Y_T>0$ for the terminal quantity on which the call
is written: $Y_T=S_T$ for a single asset and $Y_T=B_T$ for a geometric
basket. Under the chosen risk-neutral measure $\mathbb{Q}$, the
time-zero price is
\begin{align}
    \label{eq:call_price_expectation}
    C(k)=e^{-r_{\mathrm{f}}T}
       \mathbb{E}^{\mathbb{Q}}\left[(Y_T-e^k)^+\right].
\end{align}
No probability density for $\log Y_T$ is assumed. 

The model-specific input is the characteristic function
\begin{align}
    \label{eq:characteristic_function}
    \phi_Y(u)=\mathbb{E}^{\mathbb{Q}}
       \left[e^{iu\log Y_T}\right],\qquad u\in\mathbb{R}.
\end{align}
We use the same notation at complex arguments whenever the defining
expectation is absolutely integrable. The four scalar inputs are
summarized in Table~\ref{tab:fourier_model_inputs}.

\begin{table}[t]
    \centering
    \small
    \begin{tabular}{lcl}
        \hline
        \textbf{Model} & $Y_T$ & $\phi_Y(u)$ \\
        \hline
	Single-asset Black-Scholes & $S_T$ & $\phi_{\mathrm{BS}}(u;T)$ \\
	Multi-asset Black-Scholes & $B_T$ & $\phi_{\mathrm{mBS}}(u\bm{w};T)$ \\
	Single-asset Heston & $S_T$ & $\phi_{\mathrm{H}}(u;T)$ \\
	Multi-asset Heston & $B_T$ & $\phi_{\mathrm{mH}}(u\bm{w};T)$ \\
        \hline
    \end{tabular}
    \caption{Scalar inputs to the common Fourier-pricing representation.
    Both multivariate cases use geometric baskets. The multivariate
    Heston model has a common variance process.}
    \label{tab:fourier_model_inputs}
\end{table}

For both baskets, we have $\log B_T=\bm{w}^{\top}\bm{s}_T$, so the scalar
characteristic function is obtained by evaluating the joint
log-price characteristic function at $u\bm{w}$. This projection
is specific to geometric baskets and does not apply directly to
arithmetic baskets or arbitrary multivariate payoffs. In particular,
the common-variance Heston basket's variance-dependent drift is
retained in $\phi_Y$.

The call-price function $C(k)$ does not decay sufficiently rapidly as $k\to-\infty$, preventing the existence of its Fourier transform in general.
Following \citet{carrMadan1999}, we therefore introduce the exponentially damped call-price function
\begin{align*}
    c(k)=e^{\alpha k}C(k),
\end{align*}
where the damping parameter $\alpha>0$ must be chosen such that the corresponding risk-neutral moment is finite, 
\begin{align} 
    \label{eq:damping_moment_condition} 
    \mathbb{E}^{\mathbb{Q}} \left[Y_T^{\alpha+1}\right] <\infty.
\end{align} 
Every real-valued log-price has a characteristic function on the real
axis, but this alone is not sufficient for the present method. We need
to evaluate $\phi_Y$ on the shifted line $u=v-i(\alpha+1)$,
$v\in\mathbb{R}$. Condition~\eqref{eq:damping_moment_condition}
ensures that these complex expectations are finite and that the
damped call-price function is integrable. All positive moments exist
in the Black-Scholes cases. In the Heston cases, moment explosions
can restrict the admissible values of $\alpha$, depending on the model
parameters and maturity.

The Fourier transform of the damped call-price function is defined by 
\begin{align} 
    \label{eq:psi_definition} \psi(v) = \int_{-\infty}^{\infty} e^{ivk} c(k) \, \mathrm{d}k. 
\end{align}
To derive it without a density assumption, first note that for
$z>0$ and $v\in\mathbb{R}$,
\begin{align*}
    &\int_{-\infty}^{\log z}
       e^{(\alpha+iv)k}(z-e^k)\,\mathrm{d}k\\
    &\qquad=\frac{z^{\alpha+1+iv}}
       {(\alpha+iv)(\alpha+1+iv)}.
\end{align*}
At $v=0$, Tonelli's theorem and this identity give
\begin{align*}
    \int_{-\infty}^{\infty}c(k)\,\mathrm{d}k
    =\frac{e^{-r_{\mathrm{f}}T}
       \mathbb{E}^{\mathbb{Q}}[Y_T^{\alpha+1}]}
       {\alpha(\alpha+1)}<\infty.
\end{align*}
Since $c(k)\geq 0$, this also establishes absolute integrability.
Substituting \eqref{eq:call_price_expectation} into
\eqref{eq:psi_definition}, the exchange of expectation and
integration is therefore justified by Fubini's theorem. We obtain
\begin{align}
    \psi(v)
    &=\frac{e^{-r_{\mathrm{f}}T}
       \mathbb{E}^{\mathbb{Q}}[Y_T^{\alpha+1+iv}]}
       {(\alpha+iv)(\alpha+1+iv)}\nonumber\\
    &=e^{-r_{\mathrm{f}}T}
       \frac{\phi_Y\left(v-i(\alpha+1)\right)}
       {(\alpha+iv)(\alpha+1+iv)}.
       \label{eq:psi}
\end{align}
Thus the rational denominator is determined by the call payoff,
whereas the terminal distribution enters through $\phi_Y$.
No Gaussian or constant-volatility assumption has been used.

The shifted characteristic function is bounded in modulus by the
moment in \eqref{eq:damping_moment_condition}. Consequently,
$\psi(v)=\mathcal{O}(|v|^{-2})$ as $|v|\to\infty$, so $\psi$ is
also integrable. Fourier inversion recovers the continuous call-price
function,
\begin{align}
    \label{eq:C_k_integral}
    C(k) = \frac{e^{-\alpha k}}{2 \pi}
    \int_{-\infty}^{\infty} e^{-ivk} \psi(v)\, \mathrm{d}v.
\end{align}
For every admissible value of $\alpha$, the exact Fourier inversion
yields the same option price. With the appropriate $\phi_Y$ from
Table~\ref{tab:fourier_model_inputs}, Eqs.~\eqref{eq:psi}
and~\eqref{eq:C_k_integral} apply to all four pricing cases.
Their discretization into a DFT is likewise independent of the
particular model.

The discrete Fourier transform is a linear map sending the vector
$x=(x_0,\ldots,x_{M-1})\in\mathbb{C}^M$ to
$y=(y_0,\ldots,y_{M-1})\in\mathbb{C}^M$, where
\begin{align}
    \label{eq:dft}
    y_l = \frac{1}{\sqrt{M}} \sum_{j=0}^{M-1} x_j e^{-i2\pi jl/M},
    \qquad
    0\leq l<M.
\end{align}
The normalization factor $1/\sqrt{M}$ makes the transformation unitary.
In the classical setting, the sums in \eqref{eq:dft} can be evaluated
using the FFT, whose computational complexity
scales as $\mathcal{O}(M\log M)$ \citep{cooley69,frigo05}.

With an appropriate choice of discretization, the integral in \eqref{eq:C_k_integral} can be approximated by sums of the form \eqref{eq:dft}. This allows the approximate evaluation of the call option price to exploit the computational efficiency of the fast Fourier transform 
and, potentially, its quantum implementation via the inverse
quantum Fourier transform.

Following the QFT-oriented discretization of \citet{ewen2024}, we
truncate the Fourier integral symmetrically and apply a rectangular
quadrature rule chosen so that the resulting sum has the form of a
unitarily normalized DFT.
To define the discretization, we fix a step size $\Delta_v > 0$ and $N \in \mathbb{N}$ and consider 
\begin{align}
\label{eq:Ck_from_psi}
    C(k) \approx \frac{e^{-\alpha k}}{2 \pi} \sum_{j=-N}^{N-1} e^{-i \Delta_v jk} \psi(\Delta_v j) \Delta_v.
\end{align}
The input index $j$ corresponds to the discretized Fourier variable
$v$, whereas the output index $l$ corresponds to the log-strike
grid.
We choose a lower bound for the log-strike 
$k_0 \in \mathbb{R}$ and a step size $\Delta_k > 0$ and define $k_l = k_0 + \Delta_k l$ for every $0 \leq l < 2N$. To obtain the desired form of the DFT, the step sizes are chosen to satisfy the grid relation
$\Delta_v \Delta_k = \frac{\pi}{N}$. Shifting the index for $v$, we get the desired form,
\begin{align}
    \label{eq:discretization}
    C(k_l) \approx \frac{e^{-\alpha (k_0 + \Delta_k l)}}{2 \pi} \Delta_v e^{i \pi l} \sum_{j=0}^{2N-1}e^{-i 2\pi jl/(2N)} x_j,
\end{align}
with 
\begin{align}
    \label{eq:x}
    x_j = e^{-i \Delta_v j k_0} e^{i \Delta_v N k_0} \psi(\Delta_v(j -N)).
\end{align}

Since $M=2N$, the sum in \eqref{eq:discretization} is precisely
$\sqrt{2N}$ times the $l$-th component of the DFT defined in
\eqref{eq:dft}. Hence,
\begin{align}
    \label{eq:discretization_dft}
    C(k_l)
    \approx
    \frac{e^{-\alpha(k_0+\Delta_k l)}}{2\pi}
    \Delta_v e^{i\pi l}
    \sqrt{2N}\,y_l.
\end{align}

Several alternative discretization schemes are possible. In particular, the weights arising from higher-order numerical quadrature rules could be incorporated into $x_j$. With \eqref{eq:discretization}, we chose a simple discretization as our objective is to evaluate the benchmark results of Fourier methods while minimizing the influence of factors that are not intrinsic to the computational realization of the DFT.

For fixed damping and grid parameters, the model enters the input
vector \eqref{eq:x} through $\psi$, while the DFT matrix and the
reconstruction formula \eqref{eq:discretization_dft} retain the same
form. Nevertheless, admissible damping, integrand decay, and suitable
frequency cutoffs and grid spacings depend on the model parameters
and maturity. Identical numerical settings therefore need not yield
the same accuracy in all four cases.

\subsection{Worst-of Call Options}
\label{subsec:worst_of}

To consider payoffs that do not admit a one-dimensional
reduction, we study a European worst-of call on $d$ assets
in the multi-asset Black-Scholes model
\eqref{eq:multivariate_bs_dynamics}
\begin{align}
    \label{eq:worst_of_payoff}
    C_{\mathrm{wo}}(K)
    =e^{-r_{\mathrm{f}}T}
    \mathbb{E}^{\mathbb{Q}}\left[
    \max\left\{\min_{j}S_{j,T}-K, 0\right\}
    \right]
\end{align}
where the minimum is taken over all underlying assets $j=1,\ldots,d$.
Unlike geometric basket options, the payoff depends on the
joint distribution of all underlying assets and requires a
genuinely $d$-dimensional Fourier representation.

A semi-analytical reference price can be obtained in terms
of multivariate normal probabilities \citep{johnson87}.
Decomposing the payoff according to the asset attaining the
minimum and changing to the measure $\mathbb{Q}^i$ associated
with the numéraire $S_{i,t}e^{q_it}$ yields
\begin{align}
    \label{eq:johnson}
    C_{\mathrm{wo}}(K)
    &=\sum_{i=1}^{d}S_{i,0}e^{-q_iT}
    \Phi_d\left(\bm{a}^{(i)};\Gamma^{(i)}\right)
    \nonumber\\
    &\quad-Ke^{-r_{\mathrm{f}}T}
    \Phi_d\left(\bm{d}_2;R\right),
\end{align}
where $\Phi_d(\bm{x};\Gamma)$ denotes the cumulative
distribution function of a centered $d$-dimensional normal
distribution with correlation matrix $\Gamma$. The
standardized thresholds are given by
\begin{align*}
    d_{2,j}
    &=\frac{\log(S_{j,0}/K)
    +(r_{\mathrm{f}}-q_j-\sigma_j^2/2)T}
    {\sigma_j\sqrt{T}},\\
    a^{(i)}_i
    &=\frac{\log(S_{i,0}/K)
    +(r_{\mathrm{f}}-q_i+\sigma_i^2/2)T}
    {\sigma_i\sqrt{T}},\\
    a^{(i)}_j
    &=\frac{\log(S_{j,0}/S_{i,0})
    +(q_i-q_j-\sigma_{ij}^2/2)T}
    {\sigma_{ij}\sqrt{T}},
    \qquad j\neq i,
\end{align*}
where
$\sigma_{ij}^2=\sigma_i^2+\sigma_j^2-2\Sigma_{ij}$.
The corresponding correlation matrices have entries
\begin{align*}
    \Gamma^{(i)}_{ij}
    &=\frac{\Sigma_{ij}-\sigma_i^2}
    {\sigma_i\sigma_{ij}},
    \qquad j\neq i,\\
    \Gamma^{(i)}_{jk}
    &=\frac{\Sigma_{jk}-\Sigma_{ij}-\Sigma_{ik}+\sigma_i^2}
    {\sigma_{ij}\sigma_{ik}},
    \qquad j,k\neq i,\quad j\neq k,
\end{align*}
with unit diagonal entries and symmetric off-diagonal
entries. We evaluate the multivariate normal probabilities
numerically using SciPy's implementation of the Genz
quasi-Monte Carlo method.

For the Fourier representation, let
$\bm{x}=\log\bm{S}_T$ and define
$g(\bm{x})=\max\{\min_j e^{x_j}-K, 0\}$. Its generalized Fourier
transform is given by \citep{eberlein10,hurd10}
\begin{align}
    \label{eq:worst_of_transform}
    \hat g(\bm{z})
    =(-1)^{d+1}
    \frac{K^{1+i\sum_{j=1}^d z_j}}
    {\left(1+i\sum_{j=1}^d z_j\right)
    \prod_{j=1}^d(iz_j)},
\end{align}
for $\bm{z}=\bm{u}+i\bm{\eta}$, with
$\eta_j>0$ and $\sum_{j=1}^d\eta_j>1$.
Fourier inversion then yields
\begin{align}
    \label{eq:worst_of_fourier}
    C_{\mathrm{wo}}(K)
    =\frac{e^{-r_{\mathrm{f}}T}}{(2\pi)^d}
    \int_{\mathbb{R}^d}
    \hat g(\bm{u}+i\bm{\eta})
    \phi_{\mathrm{mBS}}(-\bm{u}-i\bm{\eta};T)
    \,\mathrm{d}\bm{u},
\end{align}
where $\phi_{\mathrm{mBS}}$ is the joint characteristic
function~\eqref{eq:multivariate_bs_cf}. The damping vector
$\bm{\eta}$ generalizes the scalar damping parameter used
in the one-dimensional setting.

\subsection{QFT-based Option Pricing}
The QFT-based option-pricing construction summarized in this subsection
follows \citet{ewen2024}.
We use this construction as the QFT baseline on which the
subsequent SFFT-based method is built.

\begin{figure}
    \centering
    \includegraphics[width=\linewidth]{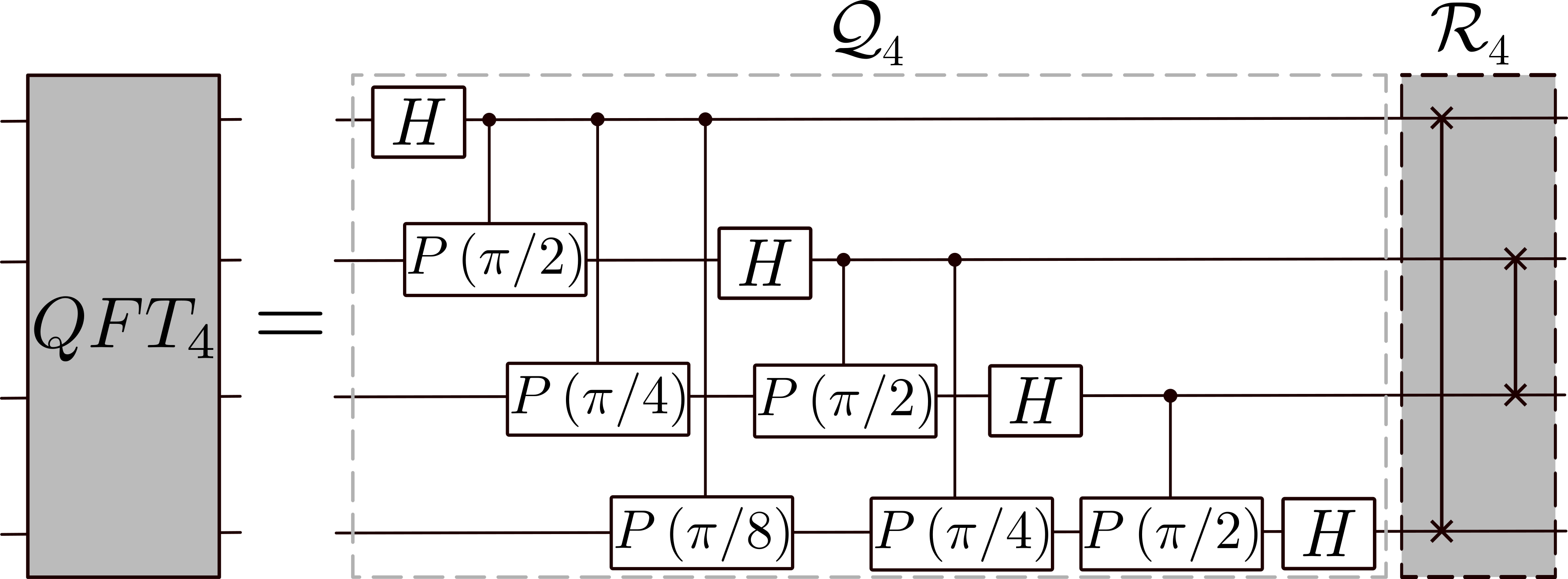}
    \caption{Quantum circuit of the QFT for the case of four qubits. The circuit separates into the actual transform $\mathcal{Q}_4$ and reordering circuit $\mathcal{R}_4$. 
    Under the sign convention adopted in \eqref{eq:dft}, the option-pricing
    calculation uses the inverse QFT, obtained by taking the adjoint of
    the displayed circuit.
    Adapted from \citet{nielsen00}.}
    \label{fig:QFT_QC}
\end{figure}

The DFT matrix in \eqref{eq:dft} is unitary for every $M$. When
$M=2^n$, it acts on a vector space that can be represented by an
$n$-qubit quantum register. Under the sign convention adopted in
\eqref{eq:dft}, the required operation is the inverse quantum Fourier
transform, since the standard QFT is commonly defined using the
opposite sign in the exponential \citep{nielsen00}. Once the
amplitude-encoded input state has been prepared, the inverse QFT can
be implemented using $\mathcal{O}(n^2)=\mathcal{O}(\log^2 M)$
elementary gates, 
see Figure~\ref{fig:QFT_QC}.

To evaluate the right-hand side of \eqref{eq:discretization} using the QFT, some technical issues must first be addressed. These arise because the amplitudes of the resulting quantum state
cannot be read out directly. Measurements in the computational basis
yield the squared magnitudes of the amplitudes, but not their complex
phases.

Recall that the total number of discretization points in
\eqref{eq:discretization} is $M=2N$. We assume that
\begin{align*}
    M=2N=2^n
\end{align*}
for some $n\in\mathbb{N}$.
To encode the vector $x$ from \eqref{eq:x} as a quantum state, it must
first be normalized with respect to the Euclidean norm. We therefore define
\begin{align*}
    \tilde{x}
    =
    \frac{x}{\lVert x\rVert}
\end{align*}
and prepare the amplitude-encoded state
\begin{align}
    \label{eq:qft_input_state}
    \ket{\tilde{x}}
    =
    \sum_{j=0}^{2N-1}
    \tilde{x}_j\ket{j},
    \qquad
    \tilde{x}_j
    =
    \frac{x_j}{\lVert x\rVert}.
\end{align}
Here, $\ket{0},\ldots,\ket{2N-1}$ denote the computational basis
states of the $n$-qubit Hilbert space $\mathbb{C}^{2^n}$.
Applying the inverse QFT to Eq.~\eqref{eq:qft_input_state} produces
\begin{align*}
    \ket{\tilde{y}}
    =
    \sum_{l=0}^{2N-1}
    \tilde{y}_l\ket{l},
\end{align*}
where
\begin{align}
    \label{eq:qft_y}
    \tilde{y}_l
    =
    \frac{1}{\lVert x\rVert\sqrt{2N}}
    \sum_{j=0}^{2N-1}
    x_j e^{-i2\pi jl/(2N)}.
\end{align}
Consequently, the sum appearing in \eqref{eq:discretization} satisfies
\begin{align*}
    \sum_{j=0}^{2N-1}
    x_j e^{-i2\pi jl/(2N)}
    =
    \lVert x\rVert\sqrt{2N}\,\tilde{y}_l.
\end{align*}
Substitution of the above relation into Eq.~\eqref{eq:discretization} therefore gives
\begin{align}
    \label{eq:discretization_y}
    C(k_l)
    \approx
    \frac{e^{-\alpha(k_0+\Delta_k l)}}{2\pi}
    \Delta_v e^{i\pi l}
    \lVert x\rVert\sqrt{2N}\,
    \tilde{y}_l.
\end{align}

The amplitudes $\tilde{y}_l$ cannot be directly obtained from measurements of the quantum state in the computational basis. Instead, such measurements yield the probabilities
\begin{align}
    \label{eq:qft_probabilities}
    p_l
    =
    \left|\tilde{y}_l\right|^2.
\end{align}
Consequently, the magnitude of the corresponding amplitude is given by
\begin{align*}
    \left|\tilde{y}_l\right|
    =
    \sqrt{p_l}.
\end{align*}

The exact call price is real and non-negative. Hence, if the Fourier
discretization is sufficiently accurate, \eqref{eq:discretization_y}
implies that
\begin{align}
    \label{eq:qft_phase_assumption}
    e^{i\pi l}\tilde{y}_l
    \approx
    \left|\tilde{y}_l\right|
    =
    \sqrt{p_l}.
\end{align}
Thus, Eq.~\eqref{eq:qft_phase_assumption} amounts to imposing the phase
expected from the positivity of the option price. 

Suppose that the quantum circuit is measured $R$ times and that the
outcome $l$ is observed $R_l$ times. The probability $p_l$ is then
estimated by
\begin{align}
    \label{eq:qft_probability_estimator}
    \widehat{p}_l
    =
    \frac{R_l}{R}.
\end{align}
Using \eqref{eq:qft_phase_assumption}, the resulting observable
approximation of the call price is
\begin{align}
    \label{eq:qft_observable_price}
    \widehat{C}(k_l)
    =
    \frac{e^{-\alpha(k_0+\Delta_k l)}}{2\pi}
    \Delta_v
    \lVert x\rVert\sqrt{2N}
    \sqrt{\widehat{p}_l}.
\end{align}

The approximation in Eq.~\eqref{eq:qft_observable_price} introduces two
sources of error beyond those arising from the discretization.
First, the approximation
\eqref{eq:qft_phase_assumption} discards any residual imaginary part or
phase error in $\tilde{y}_l$. Second, the finite number of measurements
introduces statistical sampling error in $\widehat{p}_l$. 

\section{SFFT-based Option Pricing}
\label{sec:sfft_option_pricing}
This section introduces the proposed SFFT-based option pricing algorithm. In
contrast to QFT-based approaches, which require quantum hardware to exploit the
tensor-product structure of the quantum circuit, the SFFT formulation transfers
this structure into a classical tensor network representation. Consequently,
the Fourier transformation can be performed on conventional computing hardware
while retaining the compact low-rank representation of the QFT operator.

First, we will discuss tensor trains, our tensor network geometry of choice, before we introduce the Superfast Fourier Transform (SFFT).
We will finish this section by introducing the novel SFFT-based option pricing algorithm.

\subsection{Tensor Trains}
\label{subsec:TTs}

One of the central challenges of the FFT-based option pricing approach is the
exponential growth of the underlying state space. A discretization with
$M=2^n$ points naturally leads to $n$-dimensional tensors with exponentially
many entries. Tensor network methods provide a framework to overcome this
limitation by representing high-dimensional tensors through a network of smaller interconnected
low-dimensional tensors \citep{fannes92, hackbusch09, hackbusch15}. If the
original object possesses a suitable low-rank structure, this representation
can reduce both storage requirements and computational complexity
significantly.

In this work, we employ the tensor train (TT) format
\citep{montangere18, biamonte23}, which represents an $n$-dimensional tensor as a
chain of third-order tensors. More precisely, a tensor $\mathcal{T} \in \mathbb{R}^{p^n}$ in TT format is written as
\begin{align}
\label{eq:tt_format}
\mathcal{T}(\lambda_1,\ldots,\lambda_n)
&=
A_1(\lambda_1)A_2(\lambda_2)\cdots A_n(\lambda_n).
\end{align}
The individual tensors 
$A_k\in\mathbb{R}^{r_{k-1}\times p\times r_k}$ 
are referred to as TT cores. Here, $p$ denotes the physical dimension of each mode, while 
$\{r_k\}_{k=1}^{n-1}$ denotes the set of TT ranks.
The latter determine the expressive capability of the representation and directly influence the computational cost of subsequent operations.
For simplicity, we characterize the TT structure by the maximum rank $ r=\underset{k}{\max} \; r_k $.

A related construction is obtained by extending the TT format from tensors to linear operators.
While a TT represents a tensor with one physical index per
core, a tensor train operator (TTO) represents a matrix or linear map by
assigning two physical indices to each core. These indices correspond to the input dimension $\lambda_k$ and output dimension $\mu_k$ of the operator:
\begin{align}
\label{eq:tto_format}
    \hat{\mathcal{T}}(\lambda_1,\mu_1,\ldots,\lambda_n,\mu_n)
    &=
    \hat{G}_1(\lambda_1,\mu_1)
    \cdots
    \hat{G}_n(\lambda_n,\mu_n).
\end{align}
Applying a TTO to a TT corresponds to a matrix-vector multiplication performed directly in tensor network form. This operation is central to the proposed
SFFT-based pricing algorithm, where the Fourier transform is represented as a
compressed TTO.

The main advantage of the TT representation is the reduction in storage complexity.
While a dense tensor with $n$ modes of dimension $p$ requires
$O(p^n)$ stored entries, a TT representation only requires
\begin{align*}
    O(npr^2)
\end{align*}
total elements.
Additionally, the TT formulation also provides a closed algebra for many operations required in
numerical algorithms.
Addition, element-wise or Hadamard multiplication, and operator
applications can all be performed directly on the TT cores without leaving the compressed representation. These operations may increase the ranks of the resulting TT, however, the
resulting tensors can subsequently be compressed to regain a compact
representation. 
The relative accuracy of the compressed and original TT in terms of its Frobenius norm is determined by the accuracy parameter $\varepsilon_\mathrm{T}$~\citep{oseledets11}.

For a general one-dimensional function $f(x)$ defined on the continuous
domain $x\in[0,1]$, a high-order tensor representation can readily be obtained using a suitable discretization.
This is achieved by exploiting the binary
encoding of the discretized variable
$x\approx x(\boldsymbol{\lambda})$, where
\begin{align}
\label{eq:discretized_x}
    x(\boldsymbol{\lambda})
    &\equiv x(\lambda_1,\lambda_2,\ldots,\lambda_n)
    \\
    &=
    \sum_{\ell=1}^{n} \lambda_\ell 2^{-\ell},
\end{align}
with the multi-index
$\boldsymbol{\lambda}=(\lambda_1,\lambda_2,\ldots,\lambda_n)$ and
$\lambda_\ell\in\{0,1\}$. The resulting grid is evenly spaced and given by 
\[
x(\boldsymbol{\lambda})\in
\{0,2^{-n},2\cdot 2^{-n},\ldots,1-2^{-n}\}.
\]
The discretized function can then be interpreted as an $n$-th order tensor
\begin{align*}
    f(x(\boldsymbol{\lambda}))\equiv \mathcal{T}(\lambda_1,\lambda_2,\ldots,\lambda_n)    
\end{align*}
which admits a tensor train representation with $n$ cores.

The construction of a TT representation from a single high-order tensor can be performed either by explicitly accessing all tensor entries or by exploiting the ability to evaluate the underlying function selectively. The former approach is commonly realized by
the TT-SVD algorithm, which constructs the representation through successive
singular value decompositions \citep{oseledets11}.
However, for high-dimensional
functions where individual tensor entries can be evaluated efficiently but the
full tensor cannot be stored, the TT-cross algorithm provides a more suitable alternative \citep{oseledets10cross}.
TT-cross constructs an approximation by sampling only a subset of tensor entries and is therefore well suited for black-box functions, especially when these functions exhibit sufficient smoothness \citep{nunez22}.
This sampling-based construction is used in the present work for generating the required TT representations of the option pricing quantities.

\subsection{Superfast Fourier Transform}
\label{subsec:SFFT}

The central component of the proposed algorithm is the TTO representation of the QFT, commonly referred to as the Superfast Fourier Transform (SFFT). The SFFT provides a low-rank representation of the Fourier transform that can be applied directly to TT representations while preserving their compressed structure. To construct this representation, we exploit the standard circuit decomposition of the QFT into a Fourier-transformation part and a subsequent permutation of the output indices.
As illustrated in Figure~\ref{fig:QFT_QC}, the QFT circuit can therefore be decomposed into two successive operations,
\begin{align*}
\mathcal{QFT}_n=\mathcal{R}_n\mathcal{Q}_n,
\end{align*}
where $\mathcal{Q}_n$ contains the Fourier transformation itself and $\mathcal{R}_n$ corresponds to a sequence of SWAP operations implementing the bit-reversal permutation of the output indices. This permutation restores the conventional ordering of the output indices used by the classical discrete Fourier transform.

Although the complete QFT circuit could be converted into a TTO, the additional reordering operation $\mathcal{R}_n$ destroys the locality of the tensor network and leads to prohibitively large TT ranks that scale poorly as the discretization is refined \citep{garcia21}. In contrast, the reduced QFT operator $\mathcal{Q}_n$ admits a compact representation due to the decay of correlations between increasingly distant cores, resulting in small effective ranks \citep{chen23}.
Therefore, only the corresponding low-rank representation of $\mathcal{Q}_n$ is referred to as the SFFT \citep{dolgov12}.

The circuit representation is converted into a TTO using the Zip-Up algorithm applied to $\mathcal{Q}_n$.
The procedure sequentially contracts the tensors resulting from the gates of the circuit while applying singular value decompositions and controlled truncations,
thereby yielding a TTO that directly maps the quantum circuit representation to a classical tensor train format.
Since this construction is performed only as a preprocessing step for the pricing algorithm,
the details of the Zip-Up procedure are omitted here. The interested reader is referred to \cite{stoudenmire10} for a thorough introduction.

As indicated earlier, the SFFT does not directly produce the conventional ordering of the discrete Fourier transform. Instead, it represents the Fourier transform up to the final bit-reversal permutation inherited from the quantum circuit representation. This ordering difference must therefore be taken into account when constructing operators in momentum space and is incorporated into the subsequent pricing algorithm.

The advantage of the SFFT becomes apparent when considering exponentially
large state spaces. A classical FFT applied to a grid with $M=2^n$ points
requires
\begin{align*}
O(M\log M)=O(n2^n)
\end{align*}
operations and access to all $M$ entries. In contrast, the SFFT operates directly on TT representations and exploits their low-rank structure.
The computational complexity of applying the SFFT is determined by the contraction of the SFFT-TTO with the input TT and scales as
\begin{align}
\label{eq:complexity_TTO}
O(n r_{\mathrm{SFFT}}^2 r^2),
\end{align}
where $r_{\mathrm{SFFT}}$ denotes the maximum TT rank of the SFFT operator and
$r$ the maximum rank of the input tensor train. In practical applications, the SFFT ranks remain small, typically around
$r_{\mathrm{SFFT}}\approx 10$ or smaller \citep{hauck26}, making the computational cost primarily dependent on the
compressibility of the input data.
Thus, for bounded or slowly growing TT ranks, the SFFT scales only linearly with the number of qubits $n$, providing a subexponential alternative to the classical FFT and reflecting the favorable scaling associated with the QFT.

\subsection{Pricing Algorithm}
\label{subsec:sfft_pricing}

The proposed pricing algorithm exploits the low-rank structure of tensor train representations to perform the complete pricing procedure without explicitly constructing the exponentially large tensors arising from the discretized problem. All intermediate quantities are maintained in TT format, and the required transformations and operations are performed directly on their compressed representations.
The complete algorithm involves three main tensor objects:
\begin{enumerate}
    \item the discretized exponential term given in Eq.~\eqref{eq:x},
    \item the QFT represented as a tensor train operator, and 
    \item the TT representation of the characteristic function $\psi$.
\end{enumerate}
The remaining damping and phase factor of Eq.~\eqref{eq:discretization_dft} are applied analytically during the readout of the prices, as described at the end of this section.

Firstly, the discretized exponential functions can be treated efficiently by exploiting their separable structure, which allows them to be constructed directly in TT format.
Using the definition of the discretized variable $x(\boldsymbol{\lambda})$ from Eq.~\eqref{eq:discretized_x}, a general class of discretized exponential functions can be rewritten as
\begin{align*}
g(\boldsymbol{\lambda})
&=
\exp\left(\gamma x(\boldsymbol{\lambda})\right)
\\
&=
\exp\left(\gamma\sum_{\ell=1}^{n}\lambda_\ell 2^{-\ell}\right)
\\
&=
\prod_{\ell=1}^{n}
\exp\left(\gamma \lambda_\ell 2^{-\ell}\right).
\end{align*}
The final expression is separated into individual functions of the physical
indices $\lambda_\ell$ and therefore corresponds to a rank-one TT representation. The separate cores can readily be identified as
\begin{align*}
\exp\left(\gamma \lambda_\ell 2^{-\ell}\right)
=
\begin{cases}
1, & \lambda_\ell=0,\\
\exp\left(\gamma 2^{-\ell}\right), & \lambda_\ell=1.
\end{cases}
\end{align*}
This direct construction not only avoids any approximation, but also enables highly efficient assembly of the TT.

Now, we are able to efficiently represent the $j$ dependent exponential term appearing in
Eq.~\eqref{eq:x},
\begin{align*}
\exp\left(-i \Delta_v j k_0\right)
&=
\exp\left(-i \Delta_v k_0 M x(\boldsymbol{\lambda})\right)
\\
&=
\prod_{\ell=1}^n\exp\left(-i \Delta_v k_0 M \lambda_\ell 2^{-\ell}\right)
\end{align*}
Consequently, this discretized function also admits a rank-one tensor train representation.
The multiplication required in Eq.~\eqref{eq:x} can therefore be carried out
directly in TT format using the Hadamard product \citep{oseledets11}. Since the resulting ranks are bounded by the
product of the individual ranks, multiplication with a rank-one tensor train
does not increase the ranks of the original representation.

The second required component is the SFFT operator. As described in
the last section, the reduced QFT circuit is converted into a tensor
train operator using the Zip-Up algorithm. This TTO representation can either
be constructed during the preprocessing phase or stored for repeated use. 
Since the SFFT only depends on the chosen discretization and approximation accuracy,
it can be efficiently reused for multiple pricing evaluations.
In the numerical experiments, we construct the SFFT directly.

After applying the SFFT-TTO, the resulting TT represents the vector $y$ in Eq.~\eqref{eq:discretization_dft}, up to the bit-reversal permutation of the output index inherited from the QFT circuit. This permutation is accounted for during the price readout by addressing the TT cores in reversed bit order, without explicitly applying a permutation operator.

The remaining $l$-dependent factor in Eq.~\eqref{eq:discretization_dft} is given by
\begin{align}
f(l) = \frac{\Delta_v\sqrt{2N}}{2\pi} \,
e^{-\alpha(k_0+\Delta_k l)}\,(-1)^l,
\label{eq:output_factor}
\end{align}
where we used $e^{i\pi l}=(-1)^l$. Although this factor is separable in the binary representation of $l$ and could therefore be incorporated into the TT as a rank-one tensor, we instead apply it analytically during the price readout. This choice is motivated by the potentially substantial rank increase caused by directly including the exponential damping and allows us to retain a more compact TT representation.

In practice, prices are required only on a strike window \(K_{\min}\leq K\leq K_{\max}\), corresponding to a contiguous range \(l_{\mathrm{lo}}\leq l\leq l_{\mathrm{hi}}\) of strike indices. We write
\[
l=h\,2^m+\lambda,\qquad 0\leq\lambda<2^m,
\]
and choose the smallest \(m\) for which the window intersects at most two blocks indexed by \(h\). For a fixed block \(h\), the first \(n-m\) TT-cores are evaluated at the corresponding bits of \(h\) and contracted successively, yielding a vector in the remaining virtual space. The resulting vector is then contracted with the last \(m\) cores and the block-restricted factor \(f\), which is represented as a rank-one tensor. This contraction produces the \(2^m\) prices associated with the block. Restricting the readout to the required strike window therefore avoids constructing the full vector of \(2^n\) prices. Its computational cost scales as \(\mathcal{O}(2^m r^2)\) per block, where \(r\) denotes the TT-rank.

The last component, the characteristic function $\psi$, as defined in Eq.~\eqref{eq:psi}, does not exhibit an obvious separable structure that would allow its TT representation to be constructed analytically. Therefore, its TT representation is obtained directly using the TT-cross algorithm.
Further details on its implementation are provided in the numerical experiments section.

Combining the three components yields the complete classical SFFT-based option
pricing algorithm shown in Eq.~\eqref{eq:discretization_dft}. All intermediate operations are performed in TT format, avoiding the explicit construction of any exponentially large tensors.

For the worst-of call options introduced in Section ``Worst-of Call Options'', the discretized integrand is represented as a TT with \(dn\) cores, corresponding to \(n\) qubits per asset. The cores are ordered blockwise, with each block representing the frequency variable of one asset. Exploiting the tensor-product structure of the multidimensional DFT, the corresponding SFFT operator is constructed by concatenating \(d\) one-dimensional SFFT-TTOs, each acting on its respective block. The bit-reversal permutation, as well as the damping and phase factors, are handled in the same way as in the one-dimensional case. The integrand couples the different assets through the factor \(1+i\sum_j z_j\) in Eq.~\eqref{eq:worst_of_transform} and the correlations in \(\phi_{\mathrm{mBS}}\). Consequently, its TT representation is constructed using the TT-cross algorithm over all \(dn\) cores.

\section{Experimental Evaluation}
\label{sec:experiments}
The pricing benchmarks consider European call options under the four models
introduced in Section ``Option Pricing''.
In the multi-asset cases, the payoff is a call on the
geometric basket \eqref{eq:geometric_basket}. We write $Y_0=S_0$ for a
single asset and $Y_0=B_0$ for a basket. Unless stated otherwise, the
common parameters are
\begin{align*}
Y_0 &= 100, &
r_\mathrm{f} &= 0.05, &
T &= 0.5.
\end{align*}
All individual assets have zero dividend yield. Both basket configurations
use $d=5$ assets with $S_{j,0}=100$, equal weights $w_j=1/5$, and asset-driver
correlations $R_{ii}=1$ and $R_{ij}=0.3$ for $i\ne j$.

The single-asset Black-Scholes benchmark uses $\sigma=0.3$, and the
multi-asset Black-Scholes benchmark uses $\sigma_j=0.3$ for every asset.
The effective basket parameters in \eqref{eq:bs_basket_parameters} are
\begin{align*}
\sigma_B^2 &= 0.0396, & q_B &= 0.0252.
\end{align*}
Both Black-Scholes configurations share the same analytical
reference-price calculation. At each benchmark strike, we evaluate
the Black-Scholes formula \citep{shreve2004} for the single asset and
\eqref{eq:analytic_bs_basket} for the geometric basket. The latter includes
the effective dividend correction $q_B$, although the individual dividend
yields vanish. 

Both Heston configurations use the variance-process parameters
\begin{align*}
v_0=\theta &= 0.09, & \kappa &= 2, & \xi &= 0.3.
\end{align*}
Thus, the initial and long-run variances match the individual
Black-Scholes variance $\sigma^2=0.09$. The single-asset Heston model
uses $\rho=-0.7$. The common-variance multi-asset Heston model uses
unit volatility loadings $a_j=1$ and correlations $\rho_j=-0.35$
between each asset driver and the common variance driver, in addition
to the asset correlations specified above.

For both Heston configurations, reference prices are obtained by numerical integration,
more precisely by adaptive
one-dimensional Fourier quadrature, using
$\phi_{\mathrm{H}}(u;T)$ for the single asset and the exact basket
projection $\phi_B(u;T)=\phi_{\mathrm{mH}}(u\bm{w};T)$ in
\eqref{eq:heston_basket_cf} for the multivariate model.
The integrals 
are evaluated using SciPy's
\texttt{quad\_vec}, with integration nodes chosen independently of the
benchmark Fourier frequency grid. 
The Heston reference prices are
therefore numerical approximations, rather than analytical prices or
results obtained from a finer FFT grid.

Recall, that for a register of $n$ qubits, the Fourier discretization contains $M=2^n$ grid points. Following the discretization introduced in previous sections, we use the same grid construction for all four models and define
\begin{align*}
N &= 2^{n-1}, &
\Delta k &= N^{-1/2}, &
b &= \log(Y_0)-\frac{N\Delta k}{2},
\end{align*}
and obtain the log-strike grid
\begin{align*}
k_l=b+l\Delta k,
\qquad
l=0,\ldots,2N-1.
\end{align*}
The corresponding frequency spacing is
\begin{align*}
\Delta v=\frac{\pi}{N\Delta k}.
\end{align*}
The damping parameter of the Carr-Madan representation is %
set to a value of $\alpha=2.5$.

The complete Fourier grid expands rapidly as the number of qubits increases
and contains increasingly extreme strike prices. At these extreme strikes,
the exponential factor occurring in the inverse Carr-Madan transformation
can amplify small floating-point errors, although the corresponding options
are of limited relevance for the present benchmark. We therefore evaluate
the approximation quality over the fixed moneyness interval
\begin{align*}
0.5 \leq \frac{K}{Y_0} \leq 1.5.
\end{align*}
Since $Y_0=100$ in all four configurations, this corresponds to the strike interval
\begin{align}
 K\in[50,150].
\end{align}
The interval contains in-the-money, approximately at-the-money, and
out-of-the-money options and is kept fixed for all register sizes.
Let $\mathcal{K}_n$ denote the set of strike-grid points contained in this
interval. For aggregate accuracy, we use the normalized root mean squared
error
\begin{align}
\label{eq:error_measure}
    \operatorname{NRMSE}(n,\alpha) =
    \frac{1}{Y_0} \sqrt{
        \frac{1}{|\mathcal{K}_n|}\sum_{K\in\mathcal{K}_n}
        \left[
            C(K,\alpha)
            -
            C_{\mathrm{ref}}(K)
        \right]^2
    },
\end{align}
where $C_{\mathrm{ref}}(K)$ denotes the reference price for the chosen model,
calculated as described above, and $C(K,\alpha)$ is the price produced by
the respective pricing algorithm. This error is evaluated separately
for each model. Normalization by $Y_0$ makes the error dimensionless and
permits a direct comparison across parameter settings and models.

For the strike-resolved analysis, we additionally use the normalized
absolute error
\begin{align}
    \label{eq:pointwise_error}
    e(K;n,\alpha)
    = \frac{\left|C(K;n,\alpha)-C_{\mathrm{ref}}(K)\right|}{Y_0},
\end{align}
where the dependence of the computed price on the register size $n$ is
now written explicitly. Unlike the NRMSE, this quantity does not average
over strikes and therefore reveals the variation in accuracy with
moneyness.

\subsection{Experiments using Classical Hardware and Tensor Networks}
\label{subsec:exps_tensornetworks}

The SFFT-based option pricing algorithm and its classical FFT-based counterpart are evaluated on a single node of the Beehive Cluster at Fraunhofer ITWM. The corresponding hardware specifications are provided in Table~\ref{tab:hardware_data_beehive} in the Appendix.

All execution times are measured using the \textit{process\_time\_ns()} function from Python's built-in \textit{time} module, which accounts for both user-space and kernel-space CPU time. The reported memory consumption refers to the storage required by the final numerical result: the dense array in the FFT-based approach and the collection of cores representing the resulting tensor train in the SFFT-based approach.

The only component approximated using TT-cross is the Fourier transformed damped call-price function $\psi$ in Eq.~\eqref{eq:psi}.
Its real and imaginary parts are treated separately to improve numerical stability.
Further, we employ a rank-adaptive TT-cross initialized with a random TT of rank five augmented by rank-one \emph{spikes} at the frequency $v=0$ and neighboring indices. This initialization shows improved stability, since $\psi$ is sharply localized around the origin, which can otherwise be missed by a purely random initial guess on fine discretizations. The cross approximation uses a relative stopping tolerance of $10^{-12}$.
The TT compression, performed after every TT operation, uses a relative accuracy of $\varepsilon_{\mathrm{T}}=10^{-8}$. Further details on the initialization, convergence behavior and parameter sensitivity are provided in the Appendix ``Sensitivity to the TT Hyperparameters''. In particular, the resulting pricing accuracy is governed primarily by the TT truncation tolerance, whereas variations in the cross approximation parameters have only a minor influence on the overall computational performance.

\subsubsection{Computational Efficiency and Memory Scaling}

\begin{figure*}[t]
    \centering
    \includegraphics[width=\textwidth]{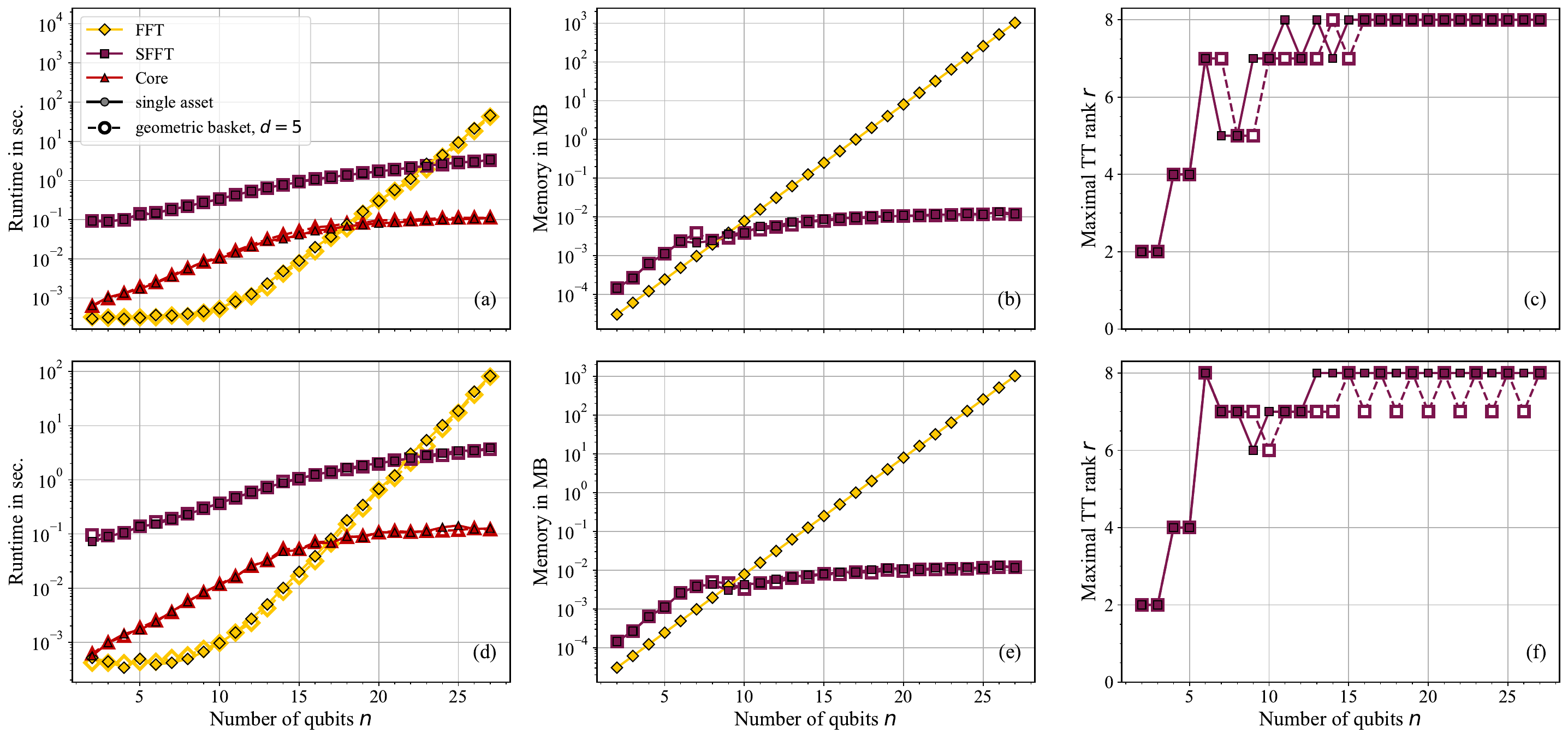}
    \caption{Classical FFT- and SFFT-based option pricing at \(T=0.5\) over the number of qubits \(n\) for the Black-Scholes models in (a) to (c) and the Heston models in (d) to (f). Filled markers with solid lines denote the single asset, hollow markers with dashed lines the geometric basket of five assets. Panels (a) and (d) show the runtime of the FFT, the complete SFFT and the SFFT Core operations, each including the readout of the strike window \([50,150]\), panels (b) and (e) the memory of the result, and panels (c) and (f) the maximal TT rank \(r\) of the result of the SFFT.}
    \label{fig:classical_overview}
\end{figure*}

Figure~\ref{fig:classical_overview} summarizes the computational efficiency and memory requirements of both approaches for the Black-Scholes and Heston models, including the single-asset and five-asset geometric-basket configurations.
The first, second and third columns show the runtime, memory consumption and maximal TT rank of the resulting representation, respectively. The corresponding accuracy results are discussed separately in the next subsection.

The memory consumption of both approaches is shown in Figure~\ref{fig:classical_overview}(b) and (e). The FFT-based approach exhibits the characteristic exponential growth in memory consumption, as it requires all grid points to be stored explicitly. In contrast, the memory requirements of the SFFT-based method increase only slightly with the discretization level, reaching approximately \(0.005\,\mathrm{MB}\) at \(n=9\) for all considered cases and remaining nearly constant thereafter. Consequently, the SFFT-based approach requires less memory than the FFT-based computations for \(n\geq9\). This behavior is explained by the bounded maximal TT ranks of the resulting representations shown in Figure~\ref{fig:classical_overview}(c) and (f). They initially increase with \(n\), but stabilize at eight for \(n\geq15\) in the Black-Scholes case and \(n\geq13\) in the Heston case. For both basket models, the maximal rank similarly settles between seven and eight for \(n\geq15\). Since the dense array has the same size for all four models and their maximal TT ranks are very similar, the corresponding memory curves essentially coincide. At \(n=27\), the SFFT representation requires only \(0.012\,\mathrm{MB}\), compared to \(1024\,\mathrm{MB}\) for the dense FFT-based array, corresponding to a reduction of almost five orders of magnitude. These results are specific to the considered problems, as the memory requirements of the SFFT-based approach generally depend on the TT ranks.

Figure~\ref{fig:classical_overview}(a) and (d) compare the runtimes of the FFT-based approach and the complete SFFT workflow. Since the construction of the components required by the SFFT can contribute substantially to its total runtime, we additionally report the runtime of the core operations, excluding these construction steps. In the following, we refer to the latter as the \textit{Core} time and to the complete workflow as the \textit{SFFT} time. Both include the readout times of the option prices over the strike window \([50,150]\).

For coarse and intermediate discretizations, the FFT-based approach remains faster. However, the Core time becomes lower than the FFT runtime at approximately \(n=18\) for Black-Scholes, while the complete SFFT workflow exhibits a crossover at approximately \(n=23\). Beyond these points, the advantage of the SFFT-based approach increases with the discretization level: the FFT-based approach runtime roughly doubles per additional qubit, whereas the Core time remains between \(0.04\) and \(0.11\,\mathrm{s}\) for \(n\geq15\). At \(n=27\), the FFT-based computation requires approximately \(46\,\mathrm{s}\), compared to \(3.4\,\mathrm{s}\) for the complete SFFT-based method and \(0.11\,\mathrm{s}\) for its Core operations.

The Heston model exhibits similar crossover points at approximately \(n=17\) for the Core and \(n=22\) for the complete SFFT-based approach. Its higher FFT-based runtime, reaching \(83\,\mathrm{s}\) at \(n=27\), is primarily due to the more expensive characteristic-function evaluations over the entire grid. In contrast, TT-cross samples only a small subset of the entries, limiting the total SFFT runtime to \(3.9\,\mathrm{s}\), while the Core time remains essentially unchanged.

The geometric-basket models show similar trends, since their payoff structure allows the pricing problem to be reduced to a one-dimensional Fourier inversion. The Core becomes faster than the FFT-based approach at approximately \(n=19\) for the Black-Scholes basket and \(n=18\) for the Heston basket. For the complete SFFT-based computation, the corresponding crossover points are approximately \(n=24\) and \(n=23\). At \(n=27\), the SFFT-based approach requires \(3.2\,\mathrm{s}\) and \(3.6\,\mathrm{s}\), respectively, compared to \(42\,\mathrm{s}\) and \(80\,\mathrm{s}\) for the FFT-based method. Thus, the observed memory and runtime advantages are also retained for these multi-asset models, provided that the payoff structure admits a dimensionality reduction.

The nearly constant Core times observed in these experiments are consistent with the stabilization of the SFFT-TTO ranks and the resulting runtime complexity predicted by Eq.~\eqref{eq:complexity_TTO}, provided that the input TT also has low ranks. More generally, the achievable efficiency therefore depends on the ranks of both the transform operator and the input tensor.

\begin{figure*}[t]
    \centering
    \includegraphics[width=\textwidth]{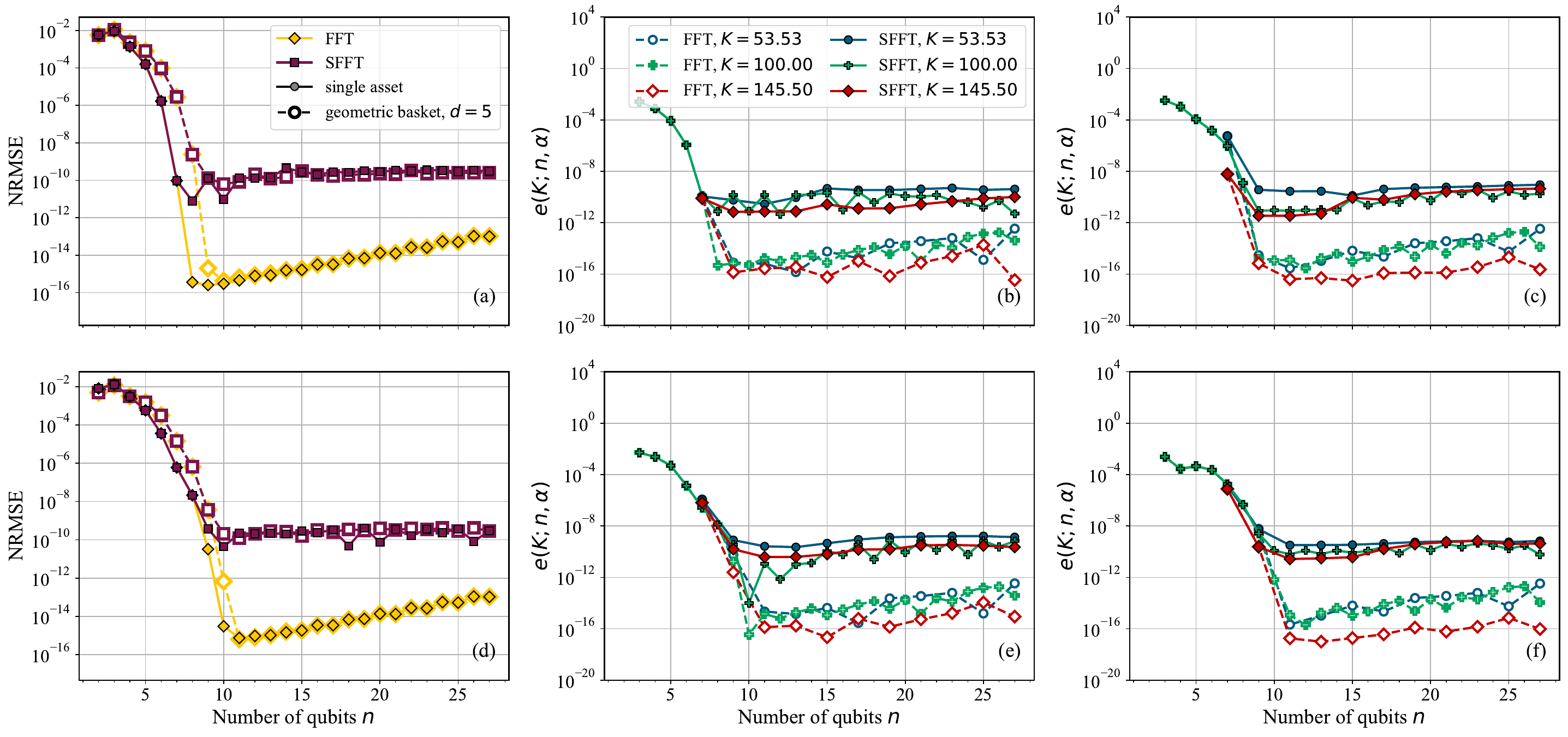}
    \caption{Accuracy of the FFT- and SFFT-based option pricing at \(T=0.5\) over the number of qubits \(n\) for the Black-Scholes models in (a) to (c) and the Heston models in (d) to (f). Panels (a) and (d) show the NRMSE~\eqref{eq:error_measure} on \([50,150]\) with respect to the closed-form price for the Black-Scholes models and an independent adaptive Fourier quadrature for the Heston models, with filled markers and solid lines for the single asset and hollow markers and dashed lines for the geometric basket of five assets. Panels (b), (c), (e) and (f) show the normalized absolute error $e(K;n,\alpha)$~\eqref{eq:pointwise_error} at three fixed strikes for the single asset in (b) and (e) and the basket in (c) and (f). Color and marker shape denote the strike, filled markers with solid lines the SFFT and open markers with dashed lines the FFT. The strikes, contracts and reference prices are the same as in the statevector simulations of Figure~\ref{fig:sim_vs_bs}, and the outer strikes are grid points only for odd \(n\).}    \label{fig:error_overview}
\end{figure*}

The runtime contributions of the individual construction steps are reported in the Appendix~``Construction Steps of the SFFT-based Approach''. At \(n=26\), the construction of the SFFT-TTO accounts for approximately \(58\%\) of the total runtime of \(3.0\,\mathrm{s}\), while the TT-cross approximation of the characteristic function contributes another \(36\%\). Importantly, the SFFT-TTO is independent of the underlying model, its parameters, the maturity, the spot price and the damping parameter, see Section~``Superfast Fourier Transform''. It can therefore be precomputed and reused across multiple pricing evaluations, for example during model calibration, across maturities of a term structure or when repricing under changing market data. If the underlying data change only moderately between evaluations, the TT-cross approximation of the input can additionally be warm-started from the previously computed approximation, which is expected to substantially reduce its runtime. At \(n=26\), precomputing the SFFT-TTO reduces the runtime per subsequent evaluation to approximately \(1.3\,\mathrm{s}\), compared to \(21\,\mathrm{s}\) for the FFT-based approach.

Further, while FFT-based implementations are highly optimized~\citep{frigo05}, the current SFFT-TTO contraction algorithm is still far from reaching its optimal implementation efficiency. Therefore, further improvements are expected through optimized contraction schemes, improved memory management and hardware-aware implementations.

\subsubsection{Numerical Accuracy and Robustness}

The NRMSE, as defined in Eq.~\eqref{eq:error_measure}, is shown in Figure~\ref{fig:error_overview}(a) and (d). For coarse discretizations, both approaches exhibit identical errors, with the accuracy limited by the discretization. Upon grid refinement, the error decreases until reaching a saturation regime. The FFT-based approach approaches machine precision, with the NRMSE increasing from \(3\times10^{-16}\) at \(n=8\) to \(1\times10^{-13}\) at \(n=27\) due to accumulated rounding errors. In contrast, the SFFT-based approach exhibits a higher error floor due to TT truncation. For the Black-Scholes model, its NRMSE stabilizes between \(2.6\times10^{-10}\) and \(3.5\times10^{-10}\) for \(n\geq17\), and at approximately \(2.5\times10^{-10}\) for the basket model. This error floor is primarily determined by the TT truncation tolerance \(\varepsilon_{\mathrm{T}}=10^{-8}\) and can be lowered by choosing a stricter tolerance at the cost of slightly larger TT ranks, as shown in the Appendix ``Sensitivity to the TT Hyperparameters''.

For the Heston models, an independent adaptive Fourier quadrature serves as reference, since no closed-form price is available. Here, discretization errors dominate up to \(n=10\), compared to \(n=8\) for the Black-Scholes models. Beyond this point, the FFT-based approach agrees with the quadrature to within rounding errors between \(6\times10^{-16}\) and \(1.1\times10^{-13}\), while the SFFT-based approach stabilizes between \(4.7\times10^{-11}\) and \(4.1\times10^{-10}\) for the single-asset model and between \(1.5\times10^{-10}\) and \(4.2\times10^{-10}\) for the basket model, for \(n\geq12\).

Figure~\ref{fig:error_overview}(b), (c), (e) and (f) complement the aggregate error measure by showing the pointwise error~\eqref{eq:pointwise_error} for the single asset and the basket at representative in-the-money, at-the-money and out-of-the-money strikes of \(K\approx53.53\), \(100\) and \(145.50\), respectively. For \(n\geq11\), the SFFT-based approach maintains errors below \(1.6\times10^{-9}\), while the FFT-based approach remains below \(3.4\times10^{-13}\) at all three strikes. The in-the-money strike typically exhibits the largest error for the SFFT-based approach, consistent with relative TT truncation.
Over the entire strike window, the maximum pointwise error of the SFFT-based approach remains less than one order of magnitude above the NRMSE. This also holds for the Heston model at the maturities \(T=0.25\), \(0.5\), \(1\) and \(2\), for which the maximum deviation from the FFT-based prices stays below \(3.5\times10^{-9}\) for \(n\geq20\). Thus, neither the strike nor the maturity introduces a pronounced deterioration in the accuracy of the SFFT.

\begin{figure*}[t]
    \centering
    \includegraphics[width=0.8\textwidth]{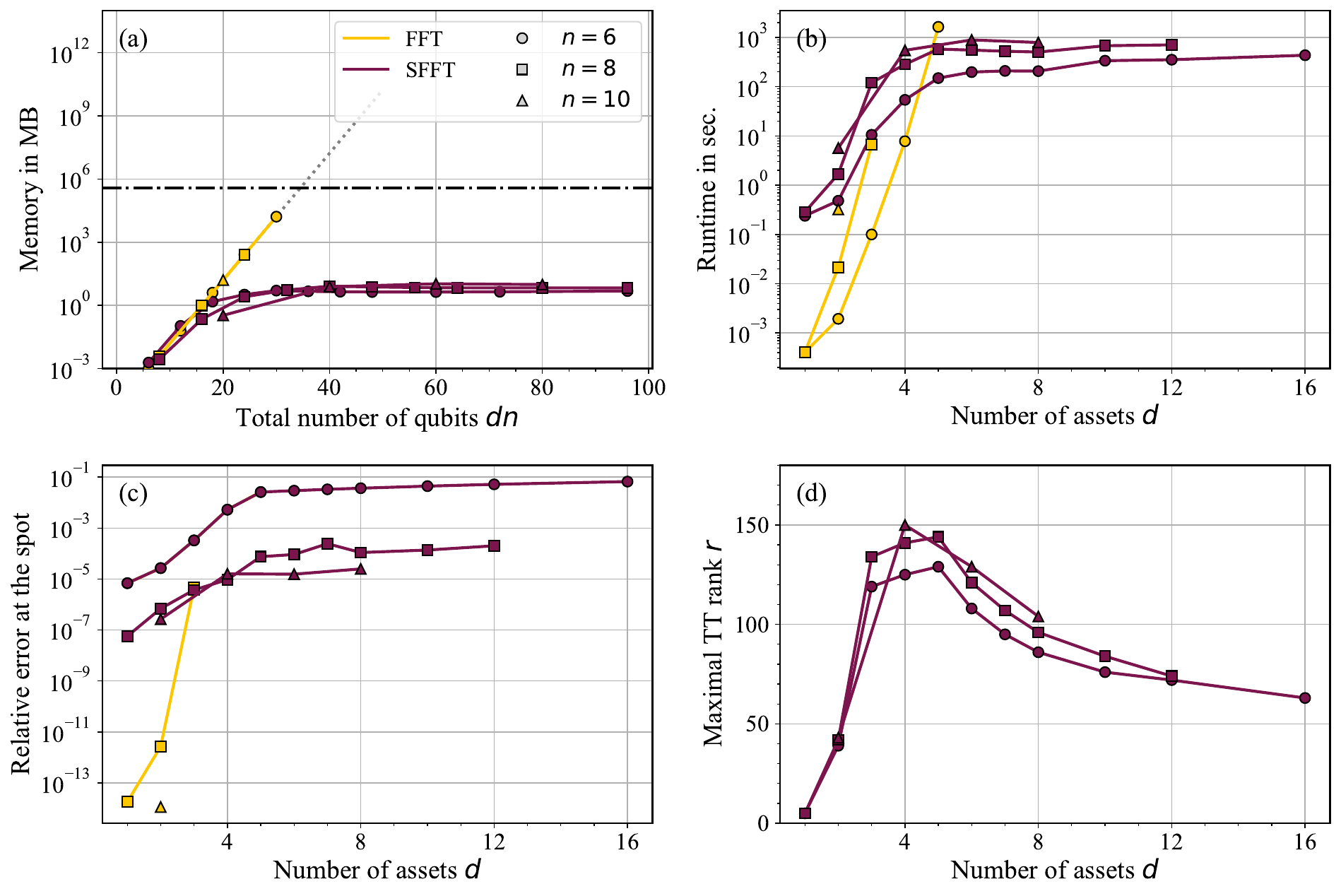}
    \caption{Worst-of call \((\min_j S_{j,T}-K)^+\) on \(d\) assets in the multi-asset Black-Scholes model, priced by a genuinely \(d\)-dimensional Fourier inversion on \(2^{dn}\) grid points. Panel (a) shows the memory of the result over the total number of qubits \(dn\), together with the memory of the dense grid
    as a dotted line and the memory of one compute node of \SI{384}{\giga\byte} as a dash-dotted line. Panels (b) to (d) show the CPU time, the relative error of the price at the spot with respect to the semi-analytical reference and the maximal TT rank \(r\) of the result of the SFFT over the number of assets \(d\). The color denotes the solver and the marker shape the number of qubits per asset \(n\). The dense FFT was only run up to $dn=30$.}
    \label{fig:rainbow_dimension}
\end{figure*}

\subsubsection{Scaling with the Number of Assets}
\label{subsec:multi_asset}
While the geometric-basket experiments reduce to one-dimensional Fourier inversions, the worst-of call introduced in Section ``Worst-of Call Options'' requires a genuinely \(d\)-dimensional inversion~\eqref{eq:worst_of_fourier} on \(2^{dn}\) grid points, with \(n\) qubits per asset. The input TT over the \(dn\) cores is constructed using TT-cross from the closed-form transformed payoff~\eqref{eq:worst_of_transform} and the joint characteristic function. The SFFT-TTO is then applied to each block of \(n\) cores.

The experiments use the multi-asset Black-Scholes model with \(S_{j,0}=100\), \(\sigma_j=0.3\) and \(q_j=0\) for all assets, equicorrelation \(R_{ij}=0.3\) for \(i\neq j\), strike \(K=100\), risk-free rate \(r_{\mathrm{f}}=0.05\) and maturity \(T=0.5\), matching the marginal parameters and correlations of the basket experiments. The damping parameter is set to \(\eta_j=\max(2.5/d,\,0.5)\), and all TT truncations and the cross approximation use a relative tolerance of \(10^{-6}\). Prices at the spot are compared with the semi-analytical reference~\eqref{eq:johnson}.

The relative pricing error at the spot in Figure~\ref{fig:rainbow_dimension}(c) is primarily driven by the discretization rather than the TT approximation. This is evident from the close agreement between both approaches wherever the FFT-based reference is feasible, with differences that are small compared with the discretization error. The discretization error, however, increases with the number of assets at fixed grid resolution. This reflects the increasingly localized structure of the multi-asset payoff, which becomes harder to resolve on a grid with an unchanged number of points per asset. For instance, at \(n=6\), the error increases from \(6.9\times10^{-6}\) for a single asset to \(6.7\times10^{-2}\) for sixteen assets.
Increasing the grid resolution substantially mitigates this effect: for \(n=8\), the error remains below \(2.5\times10^{-4}\) for up to twelve assets, while \(n=10\) achieves errors below \(2.5\times10^{-5}\) for up to eight assets. Nevertheless, the decreasing TT ranks observed at larger asset dimensions may introduce an additional source of error. Since the truncation tolerance is kept fixed across all configurations, the resulting compression becomes less restrictive in relative terms as the problem dimension increases. An adaptive truncation tolerance that accounts for the number of assets could therefore help control the overall pricing error.

The runtime of the SFFT-based approach increases with the number of assets, primarily due to the construction of the input TT. For small configurations, the FFT-based approach is faster, but its runtime grows exponentially with \(dn\), and the SFFT-based approach is faster for the largest configuration in which both methods are feasible. For \(d=5\) and \(n=6\), for example, its runtime is less than one tenth of that of the FFT-based approach, while the resulting prices differ by only \(2\times10^{-5}\) in relative terms. With \(n=8\), relative errors below \(2.5\times10^{-4}\) are obtained for up to twelve assets. Thus, the largest configurations demonstrate the memory scalability of the method, while the finer discretizations show that this scalability can be combined with accurate pricing. As in the one-dimensional experiments, the accuracy and memory requirements can be adjusted through the TT truncation tolerance \(\varepsilon_{\mathrm{T}}\).

These experiments constitute a scaling study in an idealized setting: all assets share the same volatility and initial value, and the equicorrelation structure makes the joint distribution symmetric under permutations of the assets. Realistic portfolios involve heterogeneous volatilities and non-uniform, potentially clustered correlations, which may lead to different TT ranks, particularly when strongly and weakly correlated assets are placed far apart in the core ordering. Moreover, practical products often combine worst-of features with barriers and path dependence, for which Monte Carlo methods are commonly used. A comparison with these methods is beyond the scope of this work. Investigating heterogeneous asset parameters and more general correlation structures, as well as extending the comparison to Monte Carlo methods, therefore remains an important direction for future work.

\subsection{Experiments on Quantum Hardware and Simulators}

We first investigate the execution properties of the (inverse) quantum Fourier transform on IBM quantum hardware, focusing on physical execution time and circuit complexity after accounting for the hardware's topology. We subsequently employ quantum simulators to examine the correctness and approximation quality of the quantum formulation independently of hardware noise.
We study worst-of calls only classically because encoding their multidimensional Fourier grids requires up to \(96\) qubits, well beyond the \(25\) qubits used in our largest quantum hardware benchmark.

The real-hardware experiments were performed on the IBM-Boston quantum
device, which has an IBM Heron r3
processor with 156 physical qubits. The processor
uses a heavy-hex connectivity structure with 176 physical CZ couplings. Its
native basis gates are CZ, identity, $R_z$, $\sqrt{X}$, and $X$, and the
backend control-system timestep is
\begin{align*}
\mathrm{d}t=4\,\mathrm{ns}.
\end{align*}

The benchmark circuits were constructed using Qiskit~2.3.0 and submitted
through Qiskit Runtime~0.45.1 using the second-generation Sampler primitive.
Each circuit was transpiled at optimization level~3 with respect to the
native gate set and coupling graph of IBM-Boston. No additional
error-mitigation procedure was applied.
The calibration quantities relevant to the present experiment are reported in Table~\ref{tab:hardware_data} in the Appendix.

Figure~\ref{fig:qpu_time} shows the execution time on quantum hardware as a function of the number of qubits.
The reported QPU execution times refer only to the inverse-QFT circuit and exclude the state-preparation step.
Excluding state preparation provides an optimistic quantum
baseline. This allows us to assess whether the SFFT remains competitive
when the required quantum input state is assumed to be available at no
cost.
The omission of state preparation
therefore favors the QFT-based approach and makes the comparison more
demanding for the SFFT. The observed competitiveness of the SFFT
Core runtime is thus obtained even under this favorable assumption
for the quantum method.

The plotted time is calculated from the duration of one transpiled circuit executed multiple times according to the considered number of shots.
The time values exclude queueing time, circuit construction,
transpilation and network communication.
For each executed circuit, the IBM scheduler reports a circuit
duration $d_{\mathrm{sched}}$ in units of the backend timestep $\mathrm{d}t$. The
duration (in seconds) of one physical circuit execution is therefore defined as
\begin{align*}
T_{\mathrm{exec}} = d_{\mathrm{sched}}\,\mathrm{d}t.
\end{align*}
It represents the time required for
one execution of the transpiled circuit, including the gates, delays, and
measurement operations contained in the circuit schedule.
The quantity displayed in Figure~\ref{fig:qpu_time} is therefore given as
\begin{align*}
    T_{\mathrm{total}} = T_{\mathrm{exec}} \cdot S,
\end{align*}
where $S$ is the number of shots.

\begin{figure}
    \centering
    \includegraphics[width=\linewidth]{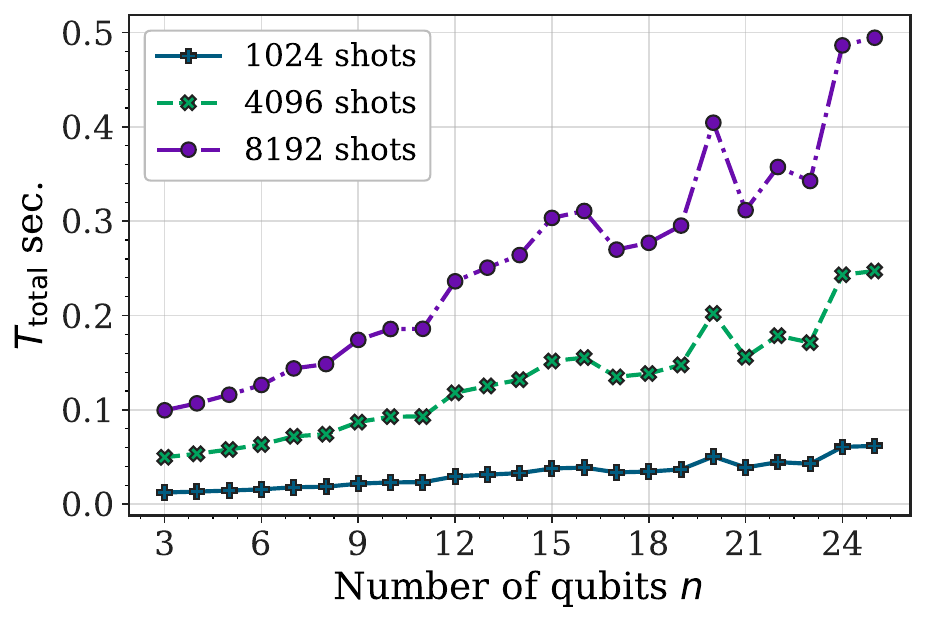}
    \caption{Runtime of quantum hardware over the discretization parameter $n$. 
    Shown are $T_{\mathrm{total}} = T_{\mathrm{exec}} \cdot S$ for different values of the
    shot parameter $S$. The default value for $S$ on IBM hardware is $S=4096$.}
    \label{fig:qpu_time}
\end{figure}

As expected, the total execution time increases with both the register size and
the number of shots. The dependence on the shot count is linear by
construction because each shot corresponds to another execution of the same
transpiled circuit.
While the represented Fourier grid grows from $8$ to $33{,}554{,}432$ points, the duration of the inverse-QFT circuit increases by only a factor of approximately five.
This behavior
reflects the polynomial growth of the inverse-QFT circuit with the number of
qubits rather than with the number of represented grid points.
The increase is not expected to be perfectly smooth. Before execution, the
logical inverse-QFT circuit is mapped to the physical coupling graph of IBM-Boston. 
The resulting circuit duration therefore depends not only on the
logical gate count but also on the selected physical-qubit layout, routing
operations, and transpiler decisions. Small fluctuations between different number of qubits are consequently attributable to compilation and scheduling
effects.

Additionally, we examine the complexity of the circuits after transpilation to the native gate set and coupling graph of IBM-Boston. 
We consider the post-transpilation circuit depth and the number of CZ gates.
Circuit depth measures the number of sequential operation layers after
parallel operations have been grouped, whereas the CZ gate count measures the
number of native two-qubit entangling operations. The latter is particularly
relevant because two-qubit operations are generally more costly and
error-prone than single-qubit operations.

The circuit statistics do not depend on the number of measurement shots.
Consequently, the configurations with multiple shots
contain repeated copies of the same transpiled circuit statistics and are
counted only once in this analysis.
\begin{figure}
    \centering
    \includegraphics[width=\linewidth]{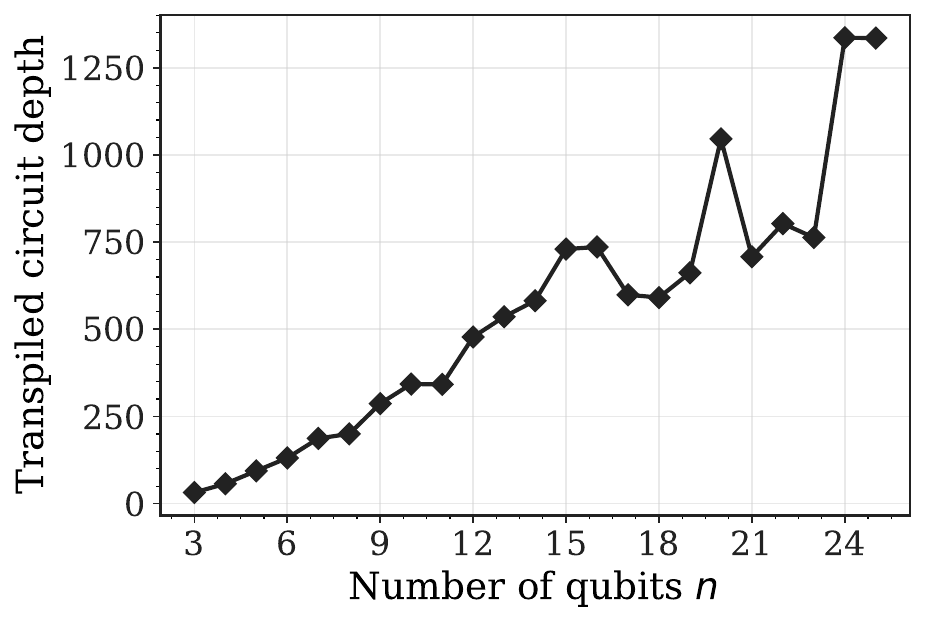}
    \caption{The depth of the transpiled circuit executed on the quantum hardware, over the discretization parameter $n$. The transpilation was performed on optimization level $3$.}
    \label{fig:depth}
\end{figure}

\begin{figure}
    \centering
    \includegraphics[width=\linewidth]{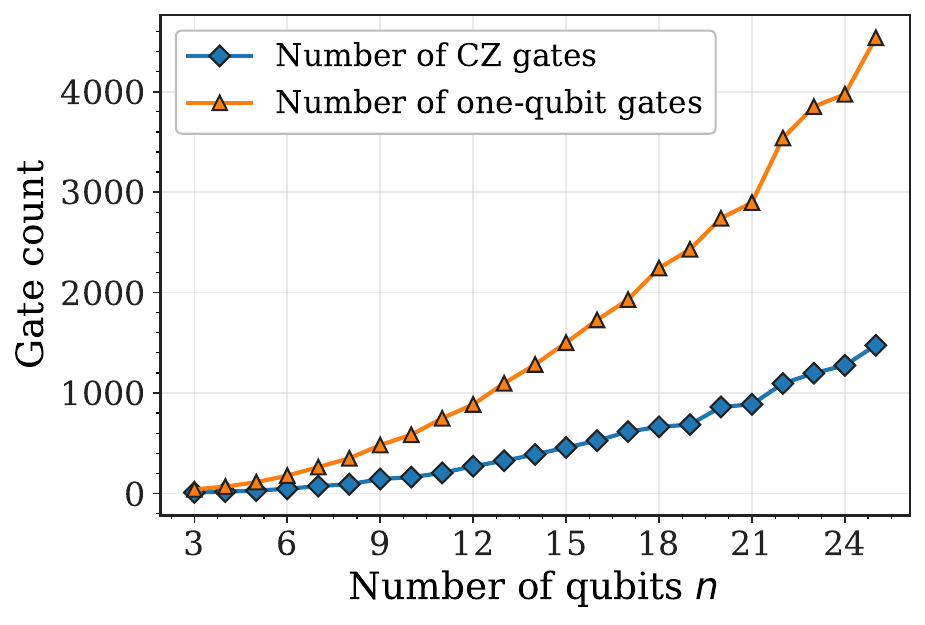}
    \caption{The number of entanglement and one-qubit gates of the transpiled circuit executed on the quantum hardware, over the discretization parameter $n$. The transpilation was performed on optimization level $3$.}
    \label{fig:gate_count}
\end{figure}

As shown in Figures~\ref{fig:depth} and~\ref{fig:gate_count}, both complexity measures increase substantially with the register size. The circuit depth increases from $32$ at $n=3$ to $1335$ at $n=25$, corresponding to an increase by a factor of approximately $40$. Over the same range, the number of CZ gates increases from $9$ to $1475$, corresponding to an increase by a factor of approximately $163$. The number of single-qubit gates also increases, from $40$ at $n=3$ to $4538$ at $n=25$.
The growth of the CZ gate count is consistent with the quadratic gate
complexity of an exact quantum Fourier transform with respect to the number
of qubits. The CZ gate count exhibits a comparatively regular increase with $n$.
Circuit depth, by contrast, is not strictly monotone.
This behavior does not indicate a reduction in the logical complexity of the
inverse QFT. Instead, circuit depth depends on how efficiently the transpiler
can place and route a particular circuit and how many native operations can
be executed in parallel. A circuit may therefore contain more CZ gates while
having a smaller depth if the selected physical layout permits greater
parallelization.

The circuit-depth results also explain the behavior of the scheduled
execution time reported above. Across the values for
$n=3,\ldots,25$, the Pearson correlation between post-transpilation depth
and time per circuit execution is approximately $0.998$.
Accordingly, the local fluctuations in the timing curve closely follow the
corresponding fluctuations in transpiled depth. 

These results emphasize the distinction between logical and physical circuit
complexity. Although the inverse QFT has a well-defined polynomial logical
gate complexity, its execution on a real processor additionally depends on
the native gate decomposition, restricted qubit connectivity, physical-qubit
placement, and routing strategy. Reporting post-transpilation complexity is
therefore necessary for interpreting the measured QPU execution times.

\begin{figure*}[t]
    \centering
    \includegraphics[width=\textwidth]{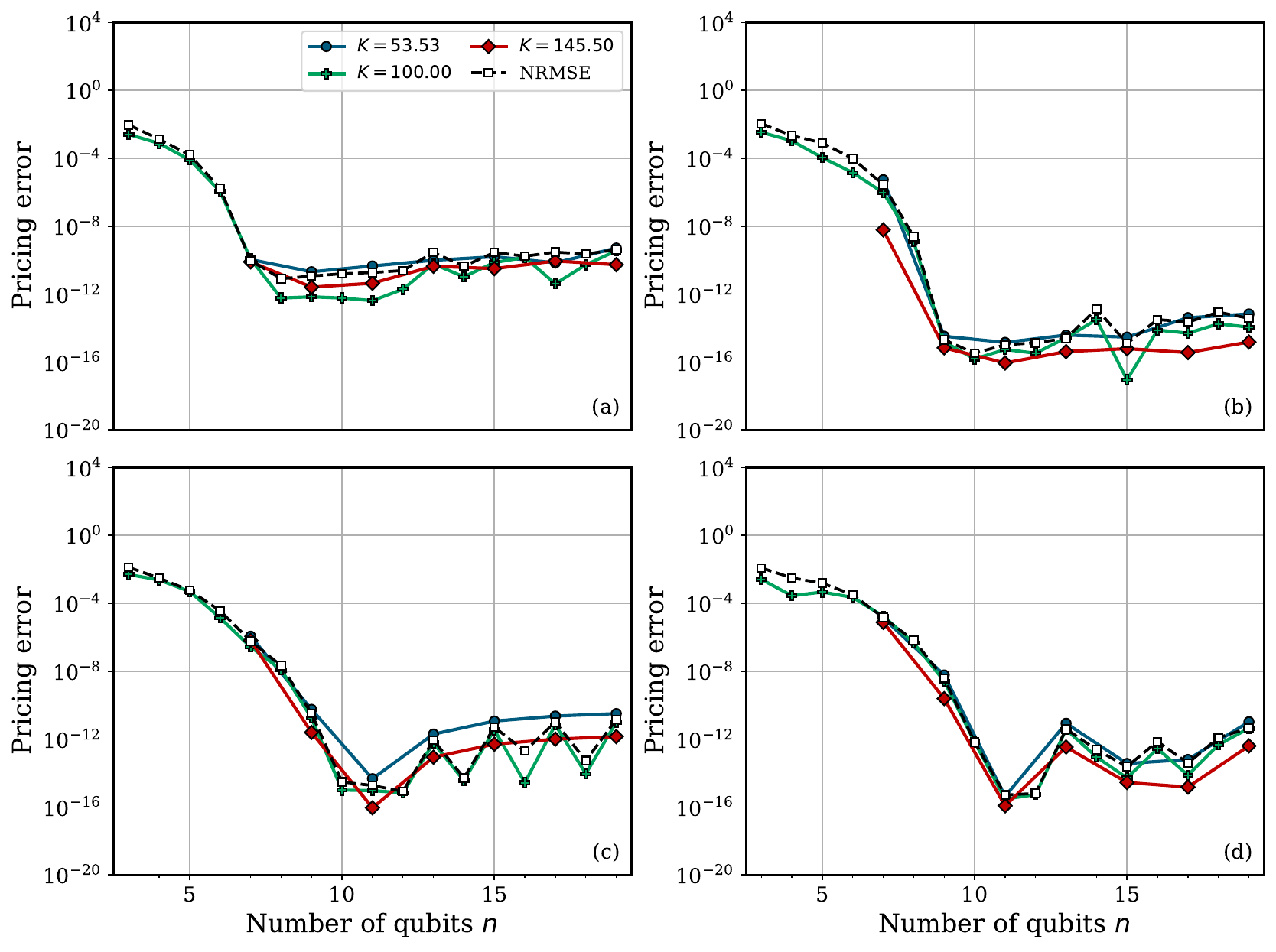}
    \caption{Strike-specific normalized absolute pricing error
    \eqref{eq:pointwise_error} for statevector simulations at
    $\alpha=2.5$: (a) single-asset Black-Scholes, (b) multi-asset
    Black-Scholes, (c) single-asset Heston, and (d) common-variance
    multi-asset Heston. Each panel contains three fixed-strike curves
    at $K\approx53.53$, $100$, and $145.50$ together with the NRMSE (dashed). The legend in (a) applies to all panels. Markers denote matching
    native grid points; connecting lines are visual guides.
    The symmetric-log vertical scale is linear below $10^{-16}$,
    retaining exact zero differences without flooring.}
    \label{fig:sim_vs_bs}
\end{figure*}

The simulator experiments are used to evaluate the numerical accuracy of the
Fourier-based pricing methods in the absence of hardware noise and sampling
error. We compare the prices obtained from statevector simulation of the
QFT-based pricing circuits with the model-specific reference prices
described above. Figure~\ref{fig:sim_vs_bs} reports the normalized absolute
error \eqref{eq:pointwise_error} for all four models at $\alpha=2.5$,
separately for the fixed strikes $K=100e^{-5/8}$, $100$, and
$100e^{3/8}$, approximately $53.53$, $100$, and $145.50$.
These represent in-the-money, at-the-money, and out-of-the-money calls,
respectively, within the chosen evaluation interval.

We use the same three strikes and strike-resolved error measure as in the classical results in Figure~\ref{fig:error_overview}(b), (c), (e) and (f), enabling a direct comparison between the quantum and classical approaches. Examining the errors across individual strikes also reveals differences that may be obscured by an aggregate error measure. For example, in the multi-asset
Black-Scholes case at $n=7$, the normalized absolute errors at the
three strikes, in increasing order of $K$, are approximately
$5.46\times10^{-6}$, $9.50\times10^{-7}$, and $6.03\times10^{-9}$.
The in-the-money error is thus almost three orders of magnitude larger
than the out-of-the-money error. The largest error need not occur at
an outer strike: for the multi-asset Heston model at $n=7$, the
at-the-money error, approximately $1.73\times10^{-5}$, exceeds the
errors at both outer strikes.

Among the evaluated grids containing all three strikes, the first
register sizes for which all three normalized errors are below
$10^{-9}$ are $n=7$ for single-asset Black-Scholes, $n=9$ for
multi-asset Black-Scholes and single-asset Heston, and $n=11$ for
multi-asset Heston. These observations apply to the selected strikes
and parameter configurations, and therefore, are not necessarily uniform error bounds over
the entire strike interval.

At finer grids, the errors fluctuate rather than decrease monotonically.
For both Heston configurations, all plotted differences at $n\geq11$
are below $10^{-10}$. However, the Heston references have a requested
absolute tolerance of $10^{-8}$ in price units, equivalent to
$10^{-10}$ after normalization by $Y_0=100$. Differences below this
scale indicate numerical agreement with the reference calculation,
but do not certify pricing accuracy at the smaller displayed values.
However, the agreement of the FFT-based approach with the same references to within \(1.1\times10^{-13}\), see Section ``Numerical Accuracy and Robustness'', indicates that their actual error is considerably smaller than the requested tolerance.

\subsection{Comparison between Classical and Quantum Methods}

\begin{figure*}[t]
    \centering
    \includegraphics[width=\textwidth]{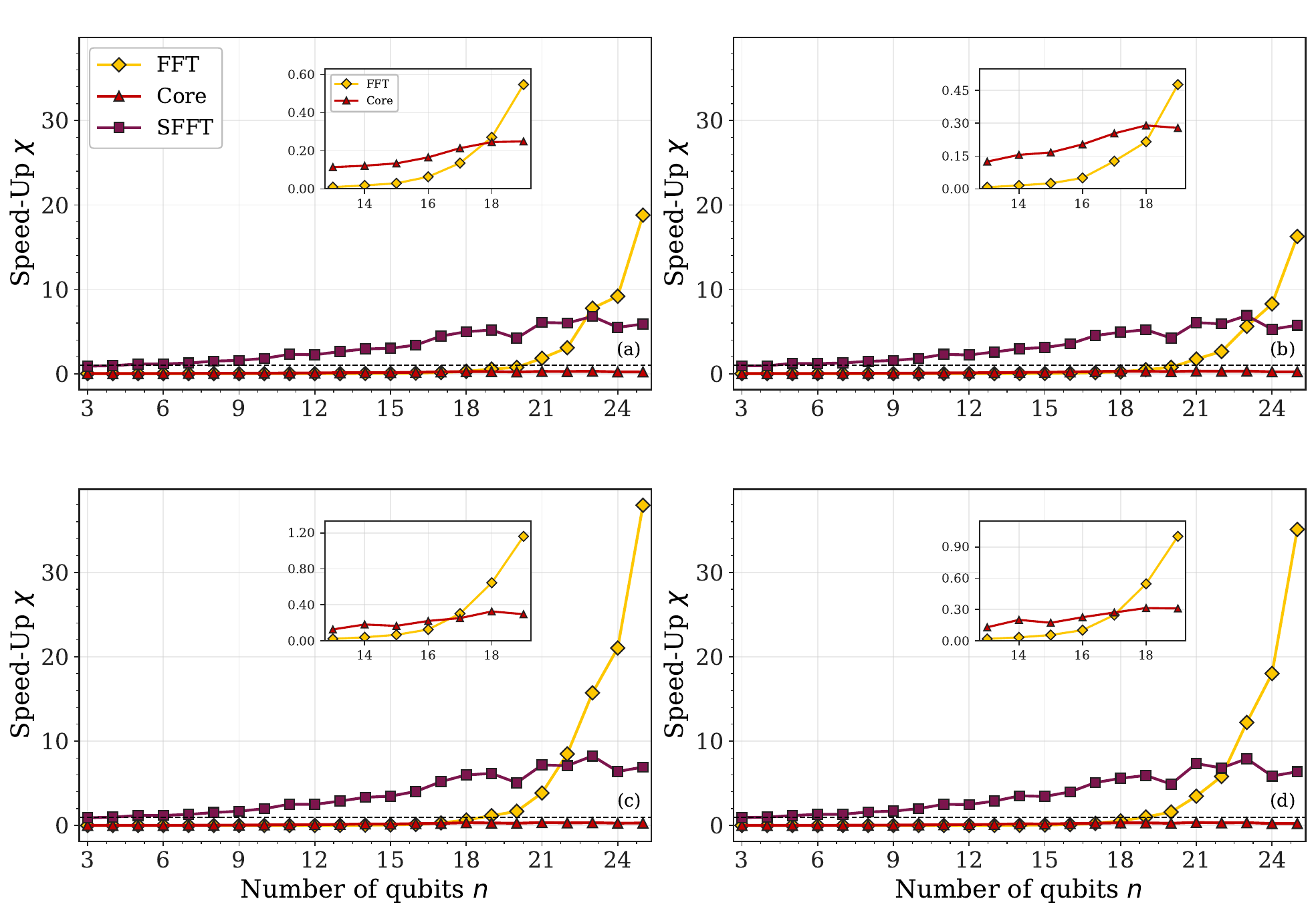}
    \caption{Speed-up measure $\chi$ comparing the runtime of the FFT-based and SFFT-based option pricing algorithms with the QPU-based implementation.
    Panels (a)-(d) show single-asset Black-Scholes, multi-asset Black-Scholes, single-asset Heston, and common-variance multi-asset Heston, respectively.
    The SFFT Core denotes the runtime of the tensor network algorithm excluding the preprocessing steps. The black dashed line at $\chi=1$ indicates the break-even point, where both approaches exhibit identical runtime.
    Each inset shows a zoomed-in view of the crossover point between the Core and FFT speed-ups. The legends in (a) apply to all panels.}
    \label{fig:speed_up_chi}
\end{figure*}

The primary objective of this section is to compare the computational efficiency of the proposed SFFT-based pricing algorithm with its classical FFT-based counterpart and the corresponding QFT-based quantum implementation.

Figure~\ref{fig:speed_up_chi} summarizes the runtime comparison between the
classical FFT-based, SFFT-based, and QFT-based option pricing algorithms for
increasing numbers of qubits $n$.
Specifically, it shows the speed-up
\begin{align*}
\chi = \frac{t_\mathrm{method}}{t_\mathrm{QPU}}    
\end{align*}
of the QPU runtime $t_\mathrm{QPU}$ with the runtime $t_\mathrm{method}$ of the respectively tested algorithm (FFT/SFFT/Core).
The QPU time is measured as the runtime of the post-transpiled quantum circuit executed on the QPU.
A value below one ($\chi<1$) indicates that the respective classical method is faster than the QPU, whereas values above one ($\chi>1$) indicate that the QPU is faster.
The dashed horizontal line at $\chi = 1$ marks the break-even point where both approaches exhibit identical runtime.

The classical FFT implementation shows the expected scaling of runtime as the discretization is refined.
For small problem sizes, the FFT-based approach remains consistently faster than the QPU execution, up to $n=20$ for the Black-Scholes models and $n=18$ for the Heston models. However, as the discretization size increases, visible around $n=18$, the exponential growth of the underlying state space dominates the computational cost.
The break-even point is crossed at approximately $n=21$ for the Black-Scholes models and $n=19$ for the Heston models, beyond which the QPU execution becomes faster. The speed-up increases rapidly with growing problem size,
indicating the exponential scaling of the classical FFT.
At $n=25$, the QPU execution time is approximately 16 to 19 times shorter than the runtime of the classical FFT-based algorithm for the Black-Scholes models and 35 to 38 times shorter for the Heston models, whose FFT-based runtime is higher due to the more expensive characteristic-function evaluations.
This trend would become even more pronounced for finer discretizations.

A different behavior is observed for the complete SFFT workflow, including SFFT-TTO construction and the associated preprocessing steps.
In this case, the speed-up starts slightly below one for small system sizes and increases approximately linearly with moderate oscillations as $n$ grows. For the largest problem sizes considered, the QPU execution is approximately six to seven times faster than the complete SFFT workflow for all four models.

As defined previously, the SFFT Core runtime isolates the computational cost of the TT-based pricing algorithm itself. Figure~\ref{fig:speed_up_chi} therefore reports this runtime separately, excluding the offline construction of the SFFT-TTO and associated preprocessing steps. In this case, the measured runtime remains nearly constant relative to the QPU across the entire range of problem sizes and is substantially smaller than both the QPU and the complete SFFT runtime, and smaller than the classical FFT runtime beyond the crossover shown in the inset. The Core runtime reaches a maximal speed-up of between $0.28$ and $0.36$ over the QPU version at a discretization of $n=21$ to $23$.
Thus, the Core runtime is at least 2.7 times faster than the QFT version for all four models.

The inset resolves the small speed-up values for the FFT and Core runtimes in the range $13 \leq n \leq 19$. While the Core speed-up remains nearly constant, the FFT speed-up increases rapidly with $n$, overtaking the Core contribution around $n=17$ to $19$, depending on the model, although both remain below the break-even value $\chi=1$ up to $n=18$.

It should be noted that the QPU runtime corresponds exclusively to the
execution of an already transpiled quantum circuit, while transpilation,
compilation, and queuing delays are excluded. Conversely, the complete SFFT
runtime includes all preprocessing costs, most notably the construction of the
SFFT representation. The comparison between the QPU execution and the SFFT
Core runtime therefore provides the most direct assessment of the respective
computational kernels. In this comparison, the SFFT Core consistently
outperforms the QPU execution across the entire range of investigated
discretizations, demonstrating the computational advantage of the tensor network implementation itself. The gap between the Core and complete SFFT
runtime further indicates that the dominant bottleneck of the current SFFT workflow is not the tensor network application, but rather the construction of
the SFFT representation.
Since this step can in principle be performed
offline and reused, for example through precomputed operator representations,
the online cost of a new evaluation reduces to the TT-cross approximation of the input and the Core operations, see Section ``Computational Efficiency and Memory Scaling''.

Comparing the statevector results in Figure~\ref{fig:sim_vs_bs} with the classical results in Figure~\ref{fig:error_overview} shows similar convergence on coarse grids, but the relative accuracy on finer grids depends on the model.
Panels~(a) and~(d) of Figure~\ref{fig:error_overview} report the NRMSE over the same strike window \(K\in[50,150]\) as Figure~\ref{fig:sim_vs_bs}, panels~(b), (c), (e) and~(f) show the normalized absolute errors at the same three strikes.
At \(n=19\), the single-asset Black-Scholes NRMSE is approximately \(4.1\times10^{-10}\) for the statevector QFT and \(3.2\times10^{-10}\) for the SFFT.
For the Black-Scholes basket, the corresponding values are \(3.7\times10^{-14}\) and \(2.0\times10^{-10}\), respectively.
At \(n=19\), the FFT gives an NRMSE of about \(7\times10^{-15}\) for both the single-asset call and the five-asset geometric-basket call.
For both Heston models at \(n=19\), the statevector NRMSE is lower than the SFFT NRMSE but also below the normalized quadrature tolerance of \(10^{-10}\), so these smaller values indicate agreement with the reference rather than verified pricing accuracy at that scale.

The classical FFT and SFFT retain the complex Fourier coefficients, whereas the quantum readout uses the probabilities \(p_l=|\tilde{y}_l|^2\).
Since the call price is non-negative, the readout approximates the phase-corrected amplitude \(e^{i\pi l}\tilde{y}_l\) by \(\sqrt{p_l}\) as in Eq.~\eqref{eq:qft_phase_assumption}; any residual phase is lost.
The statevector calculations use these probabilities without shot sampling or hardware noise.
For small \(p_l\), however, the square root can amplify absolute errors in the computed probability, so finite precision in state preparation and simulation can contribute to the fluctuations on fine grids.

\section{Conclusion}
\label{sec:conclusion}

In this work, we presented a tensor train formulation of the Carr-Madan option-pricing method based on the Superfast Fourier Transform (SFFT). By representing the QFT as a compressed tensor train operator, the proposed approach avoids explicitly constructing the exponentially large vectors required by conventional FFT-based pricing. 
In addition to the classical SFFT formulation, we implemented the corresponding QFT-based pricing algorithm on quantum hardware. This enables a direct comparison of the classical SFFT and quantum QFT formulations.

The numerical results demonstrate that the tensor network approach can substantially reduce the memory requirements of large-scale Fourier pricing while maintaining high pricing accuracy. For the considered Black-Scholes and Heston models, both for single assets and for geometric baskets of five assets, the tensor train ranks eventually saturate at about eight, leading to almost five orders of magnitude lower memory consumption than the classical FFT for fine discretizations. The SFFT also achieves comparable accuracy, with only a small increase in the final error, which is controlled by the tensor train truncation tolerance and saturates at a normalized root mean squared error of about $3\times10^{-10}$ for a relative compression accuracy of $\varepsilon_{\mathrm{T}}=10^{-8}$.
Pointwise errors at representative strikes and over the full strike window confirm that this aggregate measure is representative across strikes and maturities. While the classical FFT remains faster for small and moderate problem sizes, the SFFT-based algorithm becomes advantageous for finer discretizations, and the complete workflow is mainly limited by the preprocessing cost of constructing the compressed Fourier operator in tensor train format.
Since this operator does not depend on the model or the contract, it can be precomputed and reused, for example during calibration.

Beyond payoffs that reduce to a one-dimensional inversion, the worst-of call shows that the SFFT also applies to genuinely multidimensional Fourier inversions. Here, the method handles up to sixteen assets, corresponding to $2^{96}$ grid points, with less than $11\,\mathrm{MB}$ of memory, while the maximal tensor train rank remains at most about $150$. For these configurations, the accuracy is limited primarily by the discretization rather than the tensor train approximation, and finer grids reduce the relative error to below $2.5\times10^{-4}$ for up to twelve assets. Accuracy and memory requirements can be adjusted to the problem at hand through the grid resolution and the truncation tolerance.

The comparison between the QFT and SFFT implementations highlights their shared underlying structure. Both approaches avoid the exponential cost of explicitly representing the full Fourier transformation, despite being implemented on fundamentally different computational platforms: the SFFT on classical hardware and the QFT on quantum hardware.
In the present experiments, the SFFT operations, excluding the cost of operator construction, achieve lower computational costs than the QFT implementation, by at least a factor of about three for all four models. This holds although the measured QPU times include only the inverse QFT and assume the input state to be available at no cost.
However, when the explicit construction of the tensor train representation is included, the overall computational advantage is reduced.
While the QFT timing results were obtained on real quantum hardware, the accuracy results were evaluated using simulators, as a fault-tolerant quantum computer capable of performing the complete pricing procedure is not yet available.

Overall, the results indicate that tensor network Fourier methods provide a promising route for scaling option pricing to discretization sizes where conventional dense FFT implementations become increasingly impractical.

Future work will focus on reducing the SFFT preprocessing costs, in particular the TT-cross approximation of the input, and on optimal truncation tolerances that adapt to the number of assets. Furthermore, the multi-asset study should be extended beyond the setting considered here, in particular to heterogeneous asset parameters, general correlation structures and path-dependent products.

\section{Acknowledgment}
The authors would like to thank Tom Ewen for his continuous support during the preparation of this manuscript. This work was supported by the AMSA project funded by the state of Rhineland-Palatinate. To enhance readability and ensure comprehensiveness, portions of this manuscript were refined with the assistance of FhGenie.

\appendix

\section{Hardware Specification}
\label{app:hardware}

The classical experiments were run on a single node of the Beehive Cluster of the Fraunhofer ITWM.
The relevant parameters are provided in Table~\ref{tab:hardware_data_beehive}.

Table~\ref{tab:hardware_data} reports a selection of calibration
quantities relevant to the experiment run on the quantum computer. The relaxation time $T_1$
and dephasing time $T_2$ characterize the coherence of the physical qubits.
The CZ gate error is relevant because the transpiled inverse-QFT circuits
contain a large number of entangling gates, while the readout assignment
error characterizes the final measurement process. Since these quantities
are defined separately for individual qubits or couplings, the table reports
their median and interquartile range over the complete operational backend.

\begin{table}[b]
    \centering
    \begin{tabular}{ll}
        \hline
        \textbf{Component} & \textbf{Specification} \\
        \hline
        Cluster & BEEHIVE-MPI3 \\
        Compute nodes & 108 (Dell PowerEdge C6420) \\
        CPU & Dual Intel Xeon Gold 6240R \\
        CPU cores per node & 48 \\
        RAM per node & 384 GB (8 GB per core) \\
        Storage & 480 GB SSD \\
        \hline
    \end{tabular}
    \caption{Hardware configuration of the Beehive Cluster used for the FFT- and SFFT-based option pricing.}
    \label{tab:hardware_data_beehive}
\end{table}

\begin{table}[b]
    \centering
    \begin{tabular}{ll}
        \hline
        \textbf{Property} & \textbf{Value} \\
        \hline
        Processor & IBM Heron r3 \\
        Backend version & 1.0.4 \\
        Number of physical qubits & 156 \\
        Number of CZ couplings & 176 \\
        Backend timestep $\mathrm{d}t$ & $4\,\mathrm{ns}$ \\
        $T_1$, median [IQR]
        & $254.3,[215.7,292.1]\,\mu\mathrm{s}$ \\
        $T_2$, median [IQR]
        & $306.6,[226.2,372.4]\,\mu\mathrm{s}$ \\
        CZ gate error, median [IQR]
        & $0.148,[0.102,0.239]$ \\
        Readout error, median [IQR]
        & $0.391,[0.269,0.623]$ \\
        CZ gate duration, median
        & $68\,\mathrm{ns}$ \\
        Readout duration
        & $2.18\,\mu\mathrm{s}$ \\
        \hline
    \end{tabular}
    \caption{Hardware configuration and calibration summary for the IBM
    Boston backend. Calibration-dependent quantities are reported as the
    median with the interquartile range in brackets. CZ statistics count
    each physical coupling once.}
    \label{tab:hardware_data}
\end{table}

\section{Construction Steps of the SFFT-based Approach}
\label{app:time_breakdown}

\begin{figure*}[t]
    \centering
    \includegraphics[width=\textwidth]{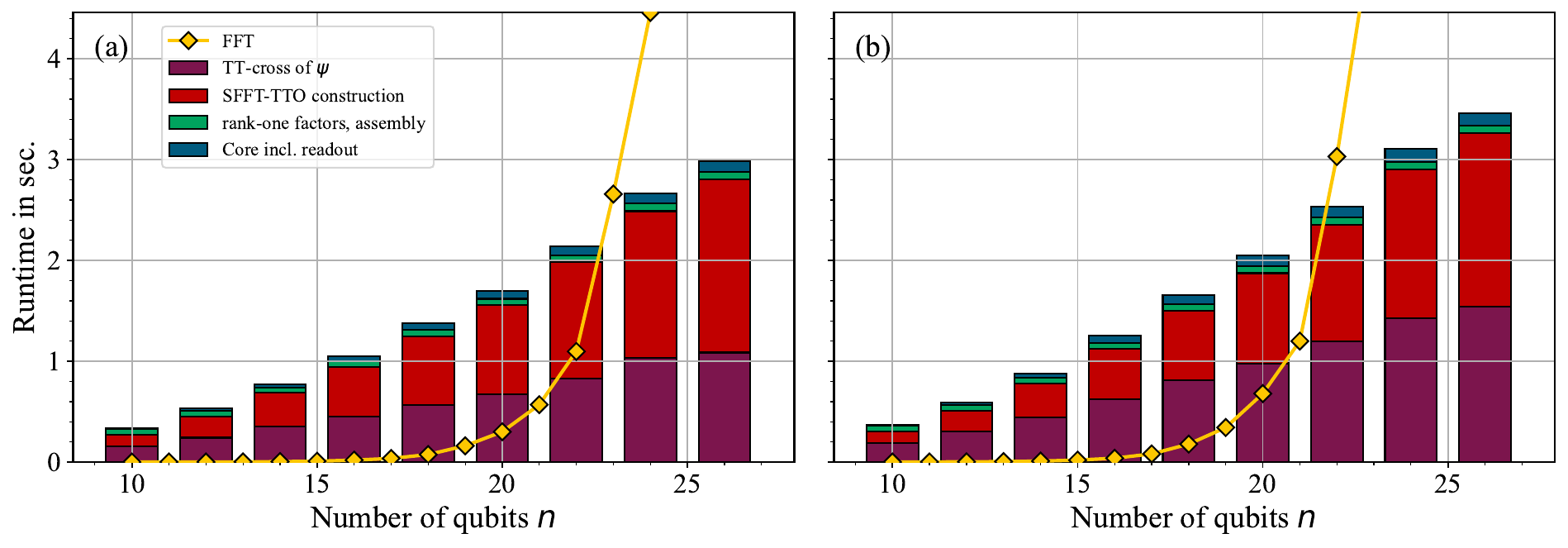}
    \caption{CPU time required by the individual construction steps of the SFFT-based approach (stacked bars) for (a) the Black-Scholes and (b) the Heston model ($T=0.5$): TT-cross approximation of the characteristic function, construction of the SFFT-TTO, rank-one factors and input assembly, and Core operations including the readout of the strike window. The line shows the FFT-based approach for comparison. The SFFT-TTO can be precomputed, whereas the TT-cross approximation and the Core operations depend on the priced contract.}
    \label{fig:sfft_time_breakdown}
\end{figure*}
\begin{figure*}[b]
    \centering
    \includegraphics[width=\textwidth]{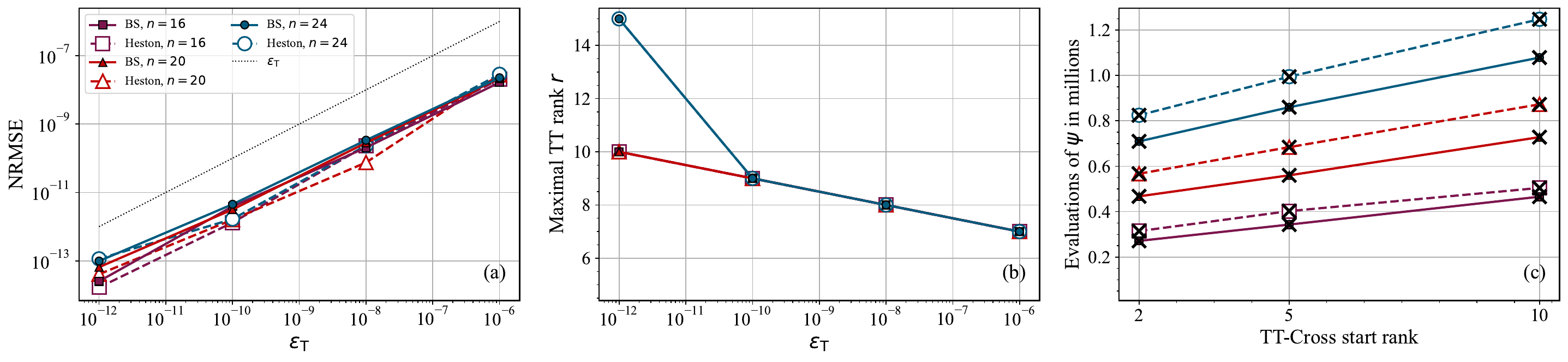}
    \caption{Sensitivity of the SFFT to its hyperparameters for $n=16$, $20$ and $24$ (colors; solid: Black-Scholes, dashed: Heston, $T=0.5$): (a) NRMSE and (b) rank of the result as functions of the TT truncation tolerance $\varepsilon_\mathrm{T}$ (cross tolerance $10^{-12}$, initial rank $5$; dotted line: $\varepsilon_\mathrm{T}$); (c) number of characteristic-function evaluations as a function of the initial cross rank, for $\varepsilon_\mathrm{T}=10^{-8}$ and cross tolerances $10^{-12}$ (lines) and $10^{-6}$ (crosses), which yield identical results.}
    \label{fig:tt_hyperparameters}
\end{figure*}

Figure~\ref{fig:sfft_time_breakdown} provides a breakdown of the CPU time required by the individual construction steps of the SFFT-based approach for the Black-Scholes and Heston models. For $n=16$-$26$, the construction of the SFFT-TTO accounts for $47$-$58\%$ of the total runtime, with the TT-cross approximation of the characteristic function accounting for most of the remaining time. At $n=26$, for example, the SFFT-TTO construction takes $1.7\,\mathrm{s}$ out of a total of $3.0\,\mathrm{s}$, while the TT-cross approximation takes $1.1\,\mathrm{s}$. The construction of the rank-one phase factors and the assembly of the input require only $0.07\,\mathrm{s}$, and the Core operations, including the readout, take $0.11\,\mathrm{s}$.

Importantly, the SFFT-TTO depends only on the discretization and the truncation tolerance, whereas the TT-cross approximation and the Core operations depend on the priced contract. The operator can therefore be precomputed and reused for multiple evaluations. With the SFFT-TTO stored, the cost of an additional evaluation at $n=26$ is approximately $1.3\,\mathrm{s}$.

\section{Sensitivity to the TT Hyperparameters}
\label{app:hyperparameters}

We investigate the influence of the TT-cross and truncation parameters introduced in Section ``Experiments using Classical Hardware and Tensor Networks'' by repeating the Black-Scholes and Heston ($T=0.5$) experiments for $n=16$, $20$ and $24$. We vary the TT truncation tolerance $\varepsilon_\mathrm{T}\in\{10^{-6},10^{-8},10^{-10},10^{-12}\}$, the TT-cross stopping tolerance between $10^{-6}$ and $10^{-12}$, and the initial cross rank between $2$ and $10$.

As shown in Figure~\ref{fig:tt_hyperparameters}, the pricing accuracy is primarily governed by $\varepsilon_\mathrm{T}$. Across both models and all tested discretizations, the NRMSE remains below $0.12\,\varepsilon_\mathrm{T}$. Tightening the truncation tolerance therefore improves the accuracy, while increasing the maximal rank only moderately: from approximately $7$ to between $10$ and $15$ at the strictest tolerance.

By comparison, neither the TT-cross stopping tolerance nor the initial cross rank has a measurable effect on the NRMSE in the tested configurations. The initial rank does influence the number of characteristic-function evaluations, but the resulting runtime variations are moderate. The stopping tolerance has virtually no effect, since TT-cross terminates after the minimum number of sweeps in all tested cases. Consequently, the overall runtime varies only slightly across the parameter combinations. These results identify $\varepsilon_\mathrm{T}$ as the primary parameter for controlling pricing accuracy, with stricter tolerances incurring only a moderate increase in computational cost.

\bibliographystyle{SageH.bst}

\bibliography{sources}

\end{document}